\documentclass[useAMS,usenatbib]{mnras}
\usepackage{graphicx}
\usepackage{graphics}
\usepackage{multicol}
\usepackage{etex}
\usepackage{setspace}
\usepackage{float}
\usepackage{pdflscape}
\usepackage{multirow}
\usepackage{hhline,booktabs,amsmath}
\usepackage{makecell}
\usepackage{caption}

\usepackage{ae,aecompl}

\usepackage{graphicx}
\usepackage{rotating}
\usepackage{epsfig}
\usepackage{amssymb}
\usepackage{amsmath}
\usepackage[utf8]{inputenc}
\usepackage{array}
\usepackage{ragged2e}
\usepackage{longtable}
\usepackage{supertabular}
\usepackage{subfloat}
\usepackage{graphicx}	
\usepackage{amsmath}	
\usepackage{amssymb}	

\title{Type-B QPOs in the black hole source H1743-322 and their association with Comptonization region and Jet}
\author[Harikrishna \& Sriram]{
S. Harikrishna,$^{1}$\thanks{E-mail:sripada.harikrishna@gmail.com}
and 
K. Sriram$^{1}$\thanks{E-mail:astrosriram@yahoo.co.in}
\\
$^{1}$Department of Astronomy, University College of Science, Osmania University, Hyderabad 500007, India\\
}

\date{Accepted XXX. Received YYY; in original form ZZZ}

\pubyear{2021}

\begin{document}
\label{firstpage}
\pagerange{\pageref{firstpage}--\pageref{lastpage}}
\maketitle

\begin{abstract}
The connection of type-B QPOs to the hot flow in the inner accretion disk region is vaguely understood in black hole X-ray binaries. We performed spectral and timing studies of twenty-three observations where type-C and type-B QPOs with similar centroid frequencies ($\sim$ 6 Hz) occurred. Their spectral differences were used to understand the production mechanism of type-B QPOs, along with the quasi-simultaneous radio observations. Based on the spectral results, we did not notice many variations in the Comptonization parameters and the inner disk radius during type-C and type-B QPOs. We found that the structure of the Comptonization region has to be different for observations associated with type-C and type-B QPOs  based on the CompTT model. Radio flux density vs QPO width, soft to hard flux ratio, and QPO width vs inner disk temperature, were found to follow certain trends, suggesting that a jet could be responsible for the type-B QPOs in H1743-322. Further studies are required to uniquely constrain this scenario. In a case study where a gradual transition from type-C to type-B QPO was noticed, we found that the spectral changes could be explained by the presence of a jet or a vertically extended optically thick Comptonization region. The geometrical Lense-Thirring precession model with a hot flow and a jet in the inner region was incorporated to explain the spectral and timing variations. 

\end{abstract}

\begin{keywords}
accretion, accretion discs---stars: individual---X-rays: binaries
\end{keywords}



\section{Introduction}
	 
	Black hole X-ray binaries (BH XRBs) exhibit low-frequency quasi-periodic oscillations (LFQPOs) typically $\leq$ 30 Hz, that are generally categorized into type-A, B, and C (Remillard et al. 2002; Casella et al. 2005). Type-A QPOs are the least common and normally appear in the soft state. These are characterized by a broad (quality factor, Q = $\nu$/$\delta$$\nu$$\leq$ 3, $\nu$ and $\delta$$\nu$ are centroid frequency and width of the QPO respectively) and weak (few percent rms) peak around 6--8 Hz. Type-C QPOs are often seen in the hard intermediate state (HIMS) and during the bright end of the low hard (LH) state  of BH XRBs along with a sub-harmonic. These are characterized by a relatively strong and narrow (Q$\geq$ 8) peak with $\sim$8--10\% rms and centroid frequencies are generally observed in the range of $\sim$1-30 Hz (e.g. Motta et al. 2011, Ingram \& Motta 2019). Type C QPOs are explained generally using a geometrical model i.e. Lense Thirring (LT) precession model (Ingram et al. 2009) or an instability model (e.g., TLM; Titarchuk \& Florito 2004).

As the BH XRB source switches from HIMS to SIMS, type-B QPOs appear in a limited range of 4--6 Hz with a low quality factor (Ingram \& Motta 2019). Most of them appear at the peak of an X-ray outburst off-setted by a radio peak (Fender et al. 2009; Motta et al. 2011). Moreover, it is noticed that their characteristic features during the flip-flop transitions are often seen in BH XRBs.  Such a rapid transition among QPOs associated with type-B was seen in GX 339-4 (Miyamoto et al. 1991; Nespoli et al. 2003; Motta et al. 2011). Gradually these rapid transitions connected to type-B QPOs were seen in other sources viz. XTE J1859+226 (Casella et al. 2004; Sriram et al. 2013), H1743-322 (Homan et al. 2005; Sriram et al. 2021), XTE J1817-330 (Sriram et al. 2012), Swift J1658.2-4242 (Bogensberger et al. 2020, Jithesh et al. 2019), MAXI J1659–152 (Yamaoka et al. 2012). Spectral studies during these events indicate that either the hard or soft component varied during the appearance of type-B QPOs. The connection of jets to type-B QPOs is becoming more  evident. 
The precession of jet is not ruled out to explain type-B QPOs (de Ruiter et al. 2019; Kylafis et al. 2020) and a geometric model potentially can explain these features, but the presence of a jet or elongated corona feature is needed (e.g. GX 339-4; Stevens \& Uttley 2016; XTE J1550-564, Sriram et al. 2016). Observationally, a relativistic jet is a noted feature during the appearance of type-B QPOs. These QPOs do not have any connection with the broadband noise component of the power density spectrum (PDS) and it is inclination independent (Arur \& Maccarone 2019, 2020), whereas Motta et al. (2015) found an inclination dependency and argued for a different origin mechanism than for type-C QPOs. Other than LT model, accretion ejection instability (AEI) can explain type-B QPOs (Varniere and Tagger, 2002, Varniere et al. 2012). 

Stella \& Vietri (1998) and Stella et al. (1999) proposed a Relativistic Precession Model (RPM) to describe the QPOs in the PDS of NS and BH XRBs. According to this model, the Lense-Thirring (LT) precession of a test particle around a black hole causes type-C QPOs in BH XRBs. The precession of hot inner flow along with a steady outer Keplerian disk can explain the evolution of type-C QPOs from 0.1-10 Hz (Fragile et al. 2007; Ingram et al. 2009; Ingram \& van der Klis 2013) where the frequency increases as the size of the hot flow shrink. The quasi-periodic modulation of the iron line in H1743-322 indicates the LT model (Ingram \& Done 2012). Motta et al. (2018) studied the precession of the test particle and the global rigid precession of a radially extended thick disk to account for the occurrence of type-C QPOs. They found the latter to be a more suitable mechanism for the production of the range of frequencies seen in type-C QPOs. They state that a radially narrow and thick accretion disk is required to sustain this global rigid precession. Moreover, a coherent QPO may not be produced if the thick accretion flow extends beyond 10-100 R$_g$. These models are successful in explaining the type-C QPOs but not often used to explain the type-B QPOs, though never ruled out (Ingram \& Motta 2019). First, such a  geometrical model was used to explain type-B QPOs in GX 339-4 where QPO is originated due to quasi-periodic illumination of the disk by a precessing Comptonizing region, probably a jet (Stevens \& Uttley 2016).

The X-ray black hole transient H1743-322 was first discovered by the
Ariel and HEAO-1 satellites (Kaluzienski \& Holt 1977) during
a bright outburst in 1977. Another bright outburst, in 2003, was
detected by the International Gamma-ray Astrophysics Laboratory and it was later 
observed by the Rossi X-ray Timing Explorer (RXTE) (Markwardt \& Swank 2003) that covered other outbursts also (e.g. Zhou et al. 2013). 
Spectral and timing features of this source show similarities with another dynamically
confirmed BH XTE J1550-564, and therefore H1743-322 was classified as a black hole candidate by McClintock et al. (2009). H1743-322 is located at a distance of $\sim$ 8.5 $kpc$ with an inclination of $\sim$75$^{\circ}$ (Steiner et al. 2012).

The production mechanism of type-C QPOs is often explained by the Lense-Thirring precession of hot flow (Ingram et al. 2009), but the type-B QPO origin is least understood. We selected type-C \& B QPOs whose central frequencies are similar such that any spectral variation between them may be connected to the type-B QPO production mechanism. It is known that type-B QPOs are connected to weak/transient jets, while type-C QPOs are connected to strong/steady jets (Fender et al. 2009). To understand this scenario, we studied 23 observations of H1743-322 among which both type-C \& B QPOs of similar frequencies are present. Quasi-simultaneous radio observations (McClintock et al. 2009) are also used to explain the association between QPOs, and jets.  

\section{Data Reduction and Analysis}
We have studied the RXTE (Swank 1999) archival data of the BH XRB H1734-322
during its outburst phase in 2003 (Markwardt \& Swank 2003, Homan 2005, Motta 2015 see Table 2). HEASOFT $v6.19$ was used for data reduction and analysis. We have selected 23 observation IDs (ObsIDs) that exhibit consistent QPOs near 6 Hz (Table 1) with radio observations reported within a time difference $\le$ 72 hr (Table 2). Single bit mode data were used to obtain the PDS with a time resolution of $\frac{1}{512}$ s in the 3.68--5.71 $keV$ 
and 6.12--14.76 keV energy bands. The obtained PDS was fitted with the Lorentzian and power-law models along with a constant (see Table 1).  

The standard 2 mode data were acquired from the Proportional Counter Array unit PCU2 (Jahoda et al. 2006). 
Spectra were obtained in the 3--25 keV energy band and a 0.5\% systematic error was added during analysis to account for 
calibration uncertainties (Sriram et al. 2021). Background spectra and responses were obtained using the PCABACKEST and PCARSP tools respectively. XSPEC $v12.9$ (Arnaud et al. 1996) was used to unfold the obtained spectra using multi-component models as discussed below. 
Spectral parameters and errors were obtained with a 90\% confidence level. We used HEXTE (Rothschild et al. 1998) data which has clusters A and B, to perform spectral analysis. We preferred HEXTE A cluster data for better sensitivity compared to cluster B. We applied hxtback command to obtain the source and background data and created the respective spectra with a rocking interval of 32 s. We applied the deadtime corrections to the obtained spectra using the hxtdead command.

For each of the ObsIDs, we present the nearest radio observations obtained from the Very Large Array (VLA) 
telescope, at 4.86 GHz and 8.46 GHz (McClintock et al. 2009) in table 2. All the observations show higher radio flux densities at 4.86 GHz compared to 8.46 GHz. A correlated X-ray and radio observational study was performed to understand the associated jet ejection mechanism. 

\section{Timing analysis}

We have obtained the PDS of light curves in two different energy bands viz. 3.68--5.71 keV and 6.12--14.76 keV. 
QPOs are consistent throughout the light curves in all the observations. The Lorentzian function is used to model the QPOs, and the obtained centroid frequencies, widths, Q, and rms values are reported in Table 1. 
 
QPOs are classified as type-C and B based on Q and rms amplitude. 
Centroid frequencies are found to be slightly higher in the higher energy band (6.1--14.76 keV) 
compared to the lower energy band (3.68--5.71 keV) in most of the observations, as reported by Li et al. (2013), except in a few observations where the centroid frequencies lie within error bars (Table 1). The width of type-C QPOs is found to be $\sim$ 0.72--1.03 Hz 
in the higher energy band while its found to be lower in the lower energy band. Similarly, the width of type-B QPOs lies in the range of $\sim$ 0.91--2.9 Hz in the 6.12--14.76 keV energy band except for two observations while that is always found to be 
comparatively higher in the 3.68--5.71 keV energy band. Q values for type-C QPOs range from $\sim$ 6.01--8.49, and for type-B QPOs range from $\sim$ 1.83--6.29 in the higher energy band while the values in the lower energy band mimic changes according to their width. 
RMS \% values in the higher energy band range from $\sim$ 9.88--15.05 for type-C 
and $\sim$ 4.69--8.75 for type-B QPOs. Sometimes a weak harmonic is present as a common feature in type-C QPOs (Table 1). 
In the case of type B QPOs, a few observations exhibit harmonic QPOs. Figure 1 shows the occurrence time of QPOs in the ASM light curve.
	
 For type-C QPOs, the full width at half maximum (FWHM) ranges from $\sim$ 0.72--1.03 Hz, and quasi-simultaneous radio flux density ranges from $\sim$ 21.49--41.44 mJy. In the case of type-B QPOs, FWHM was found to vary in the range of 0.91--2.9 Hz and radio flux densities from 0.34--21.49 mJy. For type-C QPOs, radio observations (4.86 GHz \& 8.46 GHz) were noted before ObsID 80146-01-30-00 with a time difference of $\sim$ 8.30 hr, while ObsIDs 80146-01-31-00, 80146-01-32-00 and 80146-01-03-00 are $\sim$ 1.03 hr, $\sim$ 5.54 hr \& $\sim$ 18.20 hr away from radio observations (4.86 GHz) respectively. The remaining ObsIDs have a larger time difference between the X-ray and radio observations. In the case of type-B QPOs, the nearest radio observations to ObsIDs 80146-01-51-00, 80144-01-01-02, 80144-01-01-00 \& 80146-01-16-00 are $\sim$ 2.29 hr, $\sim$ 8.38 hr, $\sim$ 8.78 hr \& $\sim$ 11.14 hr away from 8.46 GHz radio observations respectively. The remaining ObsIDs have a larger time difference. We observed an increasing trend between type-B QPO FWHM and radio flux density (Figure 2) which is unclear for type-C QPOs. In this plot, we have considered only those ObsIDs that have the nearest radio events without repeated values.

\section{Spectral analysis}   

For spectral analysis, several  combinations of spectral models were used, including the diskbb, CompTT, Power-law, SIMPL \& cutoff power-law models. We used different multi-component models to understand the spectral variations in type-B and C QPO observations. The hydrogen column density $N_{H}$ was fixed at $2.2\times 10^{22} cm^{-2}$ (McClintock et al. 2009).
A distance of 8.5 Kpc and an inclination angle of $75^{\circ}$ were used for estimating 
the inner disk radius in the analysis (Steiner et al. 2012).

PCA spectral data of sections exhibiting type-C QPOs were best fitted with using diskbb + Gaussian + CompTT  (model 1), 
assuming a spherical geometry for the Comptonization region (Figure 3 top left panel; Table 3). The iron line at 6.4 keV was fitted using a Gaussian model. Consecutive ObsIDs from 80146-01-29-00 to 80146-01-33-00, exhibit a consistent type-C QPO for nearly 4.1 days. 
Spectral fits for these ObsIDs were obtained and shown in Table 3. The inner disk temperature ($kT_{in}$) varied in a range of $\sim$ 0.81--1.07 $keV$, while the electron temperature (kT$_{e}$) varied from $\sim$ 5.87--10.19 keV. Initially, the input soft photon temperature was allowed to vary but later needed to be fixed to constrain the error bars. We noted a significant difference between the soft (diskbb) and hard (CompTT) component flux values (see Table 3). The true inner disk radius, 1.2$\times$(N$_{diskbb}$/cos $\theta$)$^{1/2}$ $\times$ (D/10 kpc) km  (Reynolds \& Miller 2013), was found to be range from 40--55 km.\\

 Soft seed photon temperature was allowed to vary and later fixed. We assumed another scenario where soft seed photons arrive from the inner disk region i.e. kT$_{o}$ = kT$_{in}$. It was noticed that R$_{in}$ values were similar where disk and CompTT fluxes were low in ObsIDs 80146-01-02-00, 80146-01-31-00, and 80146-01-03-00, and for the remaining observations R$_{in}$ significantly changed when compared to values where kT$_{0}$$\neq$kT$_{in}$. The inner disk temperature was found to be low compared to the values using model 1 (Table 3).

The same model was used to unfold the spectra connected to the type-B QPO observations, and we noticed that it resulted in an unrealistically high kT$_{e}$ (e.g., for ObsID 80146-01-62-00, we found that kT$_{e}$ $\sim$ 87 keV and unable to constrain). Hence we tried the $wabs(simpl\times diskbb + power law + Gaussian)$ model (model 2; Table 4). The SIMPL model takes care of the diverging power-law component connected to the soft energy domain (Steiner et al. 2009). Here, the power-law index ($\Gamma$) was needed to arrive at a reasonable $\chi^{2}$/dof (Shaw et al. 2016). Soft seed photons were allowed to up-scatter during the fit. $kT_{in}$ was found to vary from $\sim$ 1.15--1.22 $keV$, and the SIMPL power-law index was ranging from $\sim$ 2.14--3.59. True inner disk radius and power-law indices are mentioned in Table 4. We also attempted to fit the spectra associated with type-C QPOs using model 2, but the data didn't demand it.

ObsIDs 80146-01-11-00 and 80146-01-12-00 are $\sim$ 6 hours apart with similar kT$_{in}$ values of $\sim$ 1.20 keV, 
but their inner radius, total flux, and component fluxes are found to be different.
A Gaussian feature at $\sim$ 6.4 keV was needed for the best fit, except in ObsIDs 80146-01-12-00, 80135-02-02-00 and 80146-01-68-00. 
For ObsID 80146-01-48-00, power-law index was not required in the best fit model.

	
For ObsIDs exhibiting type-C QPOs, inner disk temperature was found to be consistent at $\sim$ 1.0 keV and cutoff power-law index from 1.71$\pm$0.18 to 2.26$\pm$0.10, using the $wabs(diskkbb + cutoffpl + Gaussian)$ model (model 3). While for sections exhibiting type-B QPOs, $kT_{in}$ ranges from $\sim$ 1.16 to 1.29 keV and cutoff power-law index from 2.29$\pm$0.12 to 2.54$\pm$0.10. 
The inner disk radius remained almost in the same range in both cases, except type-C QPO observations with low flux. Similar to the other models, flux difference is high in type-C QPO observations and low in type-B QPO observations. A Gaussian model is found to be necessary while fitting type-C QPO observations, while it is not required for type-B QPO observations.

Apart from the ObsIDs mentioned above, We studied QPOs in ObsID 80146-01-47-00 (a star marker in Fig. 1). In the ASM light curve, this ObsID is seen to be lying between a dip to rise/high transition (Figure 1). This observation is associated with a type-C QPO (Fig. 9), 
which eventually appears as a type-B QPO in the subsequent ObsID 80146-01-48-00.

 Since the higher energy band is more appropriate to constrain the electron temperature using the CompTT model, we used HEXTE data along with PCA. We performed PCA+HEXTE spectral analysis (3--100 keV) for type-C and type-B QPO observations (Table 5). We used model 1 to unfold the spectra of type-C QPOs (Figure 3 bottom right panel). During the fits we observed that kT$_{0}$ was not getting constrained and hence tied this parameter to the inner disk temperature, assuming that soft photons from the disk are the input seed photons for the Compton cloud. We noted that for the selected type-C QPOs, kT$_{in}$ varied from 0.67--0.78 keV, along with the inner disk radius (see Table 5). kT$_{e}$ was found to vary from $\sim$ 16.12--26.96 keV with low optical depth values. Other than these two parameters, relative changes remained the same compared to PCA spectral parameters. For ObsIDs exhibiting type-B QPOs, the kTe (resulted in very high values) and other parameters of the CompTT model were not getting constrained with the same model. Moreover, we tried the Nthcomp model instead of CompTT, still could not constrain the electron temperature.
We attempted to fit spectra of type-B QPOs using model 2 (i.e. wabs(simpl*diskbb+Gaussian+power-law)), and the best fits of the spectra are shown in Table 5. kT$_{in}$, R$_{in}$ \& f$_{sca}$ values did not change but $\Gamma$ was found to vary in a few observations, eventhough type-B QPO centroid frequencies were the same. We noticed that $\Gamma_{simpl}$ was softer than $\Gamma$ in these observations.

The three sections (a), (b), and (c) of ObsID 80146-01-47-00 (Figure 9) are best fitted using model 1, with kT$_{0}$ = kT$_{in}$, model 2 (without power-law index), and model 3 (see table 6) and the obtained parameters were compared with ObsID 80146-01-48-00 (section d).
 The CompTT model parameters were constrained in ObsID 80146-01-48-00 where a type-B QPO was present (we noted a non-physical kT$_{e}$ value for the kT$_{0}$ = kT$_{in}$ case, as mentioned above). kT$_{e}$ was consistent at around $\sim$ 10 keV with $\tau$ $\sim$ 4.0, however, kT$_{in}$ and R$_{in}$ values were found to be different between c and d sections. The diskbb and CompTT fluxes did not vary much across a to c sections. A similar trend was noticed when the condition of kT$_{0}$ = kT$_{in}$ was applied.  
In model 2, $\Gamma_{simpl}$ softened from c to d sections ($\sim$2.65 to $\sim$2.76). kT$_{in}$ was found to increase ($\sim$0.88 keV to $\sim$1.10 keV), while R$_{in}$ decreased. Similar changes were noted when spectra were unfolded with model 3.

We unfolded the PCA + HEXTE spectra for a, c sections of ObsID 80146-01-47-00 and d section of ObsID 80146-01-48-00 (Figure 3, middle and bottom left panels; see Table 7). We did not consider section b as HEXTE spectrum quality was not good due to the small time duration. We used model 1, model 2, and model 4 i.e. only power-law for fitting the HEXTE spectra. Based on model 1 fits, we noted that kT$_{e}$ was found to be $\sim$ 31 keV and $\sim$ 40 keV for a and c sections whereas it could not be constrained in the d section (Table 7). kT$_{in}$ was found to be $\sim$ 0.92 keV, and almost similar values were observed in PCA spectral fits. We used model 2 instead of CompTT in section d and obtained $\Gamma$ =2.64$\pm$0.04. The addition of a power-law model in a and c sections did not affect the fits as $\chi^2$/dof did not vary significantly. The disk flux was found to increase, and similar variation was noted during the PCA spectral analysis. In model 2, a power-law is required which improved the fits probably due to the inclusion of HEXTE spectral data. $\Gamma_{simpl}$ softened from c ($\sim$ 2.60) to d section ($\sim$ 2.77) and such $\Gamma_{simpl}$ was also observed in other type-B associated observations (Table 5). It was also noted that the scattering fraction and $\Gamma$ decreased from 0.39 to 0.16 and $\sim$ 2.82 to $\sim$ 2.64 respectively. kT$_{in}$ was found to increase from $\sim$ 0.99 keV to $\sim$ 1.18 keV from c to d sections, whereas R$_{in}$ slightly decreased. We also fitted a power-law model alone to HEXTE spectra (15--100 keV) and noted that $\Gamma$ varied from $\sim$ 2.82 to $\sim$ 2.66 from c to d sections. The variation of $\Gamma$ from model 2 suggests that the emission from the Comptonization region became harder during this episode.

\section{Results and Discussion}

We have analyzed observations exhibiting type-C and type-B QPOs near 6 Hz 
that are consistent throughout the light curves of BH source H1743-322. 
The idea of this study is to know the spectral differences between type-C and type-B QPO observations that can provide information about the latter.     
RMS \% values are higher in the higher energy band and lower in the lower energy band (Belloni et al. 2020). 
On comparing the timing analysis of type-C \& type-B QPOs in the higher energy band 6.12--12.69 $keV$ (Table 1), we note that the width of the type-C QPOs is ranging $\sim$ 0.72--1.03 Hz and that for type-B QPOs $\sim$ 0.91--2.90 Hz (except in two observations). RMS\% 
is always higher for type-C QPOs. 

The nearest radio flux density values for type-C and type-B QPO X-ray observations are reported in Table 2. All these observations show higher flux densities in the lower radio band (4.86 GHz) compared to the higher band (8.46 GHz). Radio flux densities of sections exhibiting type-C QPOs are high, while they are relatively low in the case of type-B QPOs (Table 2). Based on our study of $\sim$ 6 Hz QPOs, we see that type-B QPO width increases along with an increase in the radio flux density, which is not seen for type-C QPOs (Figure 2). This increasing trend is fitted linearly with the slope of 8.35$\pm$2.80 at a 95\% confidence level, and an $F$-test value of 18.69 with a probability of 0.83$e$-3 compared to a constant fit. Though the X-ray and radio observations were not simultaneous, we argue for a scenario where QPOs and radio-jets are interrelated. Simultaneous observations would help us understand the connection better. The relation between radio and X-ray observations (Table 2, last column) can, in principle, be interpolated to know the radio emission during the corresponding X-ray observations. However, as argued by Fender et al. (2009), the one-to-one connection between radio and X-ray properties is difficult and probably inherent to the BHXBs.

In general, QPOs have been known to change their Lorentzian nature on short time scales. In some cases, it is so rapid that transitions occur within a few tens of seconds (e.g., Nespoli et al. 2003; Sriram et al., 2013, 2016, 2021). Interestingly, the series of observations from ObsID 80146-01-29-00 to 80146-01-33-00, exhibit a consistent type-C QPO for $\sim$ 4.1 days. The CompTT parameters indicate that this region was low in electron temperature and optically thick. R$_{in}$ was found to be around $\sim$ 40-55 km (53--107 km for CompTT model with kT$_{0}$ = kT$_{in}$)  during these observations (Table 3). Between ObsIDs 80146-01-29-00 and 80146-01-30-00, a decrease in disk and coronal fluxes is evident with a radio observation just before ($\sim$ 8.30 hr) 80146-01-30-00. In 80146-01-31-00, the lowest X-ray flux is noted $\sim$ 1.03 hr after the highest radio flux density (34.59$\pm$0.24 mJy) was recorded in this series of observations, which indicates a possible association. Based on the spectral analysis of type-C QPO observations, we see that ObsIDs 80146-01-02-00, 80146-01-03-00 and 80146-01-31-00 show similar trends in spectral parameters with lower flux values compared to the rest. Similar to 80146-01-31-00, $\sim$ 18.20 hrs before the observation of ObsID 80146-01-03-00, a radio emission (41.44$\pm$0.24 mJy) was observed (Table 2). We can not rule out the possibility that there could have been a similar event for ObsID 80146-01-02-00, which was not recorded.

For type-B QPO observations, model 2 resulted in an inner disk temperature range of $\sim$ 1.15–1.22 keV with a dominant disk flux component (see Table 4) and is associated with a scattering fraction of $\sim$ 0.03–0.20. Even though the $\Gamma_{simpl}$ changed from $\sim$ 2.14--3.36, it doesn't show any direct relation with the inner disk radius and QPO frequency. Power-law indices in models 2 and 3 are found to be similar within error bars. Nearest radio observations were reported at $\sim$ 2.29 hr and $\sim$ 5.21 hr before the type-B QPO ObsIDs 80146-01-51-00 and 80146-01-62-00 respectively. Interestingly, obsIDs 80146-01-11-00, and 80146-01-12-00 are consecutive observations ($\sim$ 6 hr apart) that are similar, but significant variation was seen in flux and inner disk radius values. Power-law indices in both the models have become harder from $\Gamma$ $\sim$ 2.80 to 2.68 in model 2, and 2.54 to 2.39 in model 3. A radio event is observed after ObsID 80146-01-12-00, followed by a type-C QPO transition (ObsID 80146-01-13-00; Sriram et al. 2021). Noticeable changes in soft and hard flux indicate that some physical changes might have occurred, and this cannot be due to an increase in the mass accretion rate, because if that would have been the case then the hard flux should decrease, but we noted an increase in the hard flux (e.g., in model 2, disk flux varied $\sim$ 2.92$\times$ 10$^{-9}$ erg cm$^{-2}$ s$^{-1}$ to 5.12$\times$ 10$^{-9}$ erg cm$^{-2}$ s$^{-1}$). The best fit of model 3 ObsID 80146-01-11-00 shows an inner disk radius of $\sim$ 21.02 Km, which is almost near the event horizon. For a 7.3 solar-mass black hole (Steiner et. al. 2012), the inner disk radius is $\sim$ 1.9 r$_g$, which means that r$_{H}$(radius of the event horizon) must be less than this value for a spin a$_{*}<$0.5 (Bambi et. al. 2018). This spin range is consistent with that reported in previous works. For a 11.2 M$_{\odot}$ black hole (Molla et. al. 2016), the inner radius is $\sim$1.2 r$_g$, which means r$_H$ must be less than this value for a spin a$_{*}>$0.95, which is inconsistent with current studies. For precise estimations, magnetic and frame-dragging effects should be involved. Moreover, during these observations, almost no change is observed in the inner disk temperatures (see Table 4). The increase in hard flux could be due to an increase in the vertical height of the corona, which is a possible scenario during these observations (Wang et al. 2021). 

PCA+HEXTE spectral fits for type-B QPO observations indicate that $\Gamma_{simpl}$ was softer compared to PCA spectral fits. Moreover, the R$_{in}$ values remained similar even after the inclusion of HEXTE data (Table 5). The inner disk front is truncated relatively far in most of the type-C QPO observations, but few observations have similar inner disk radii as type-B QPO observations. Our study is to strengthen the relative changes and not to constrain the absolute values of the disk and Comptonization parameters as they are specific to the models one uses. Based on our analysis we argue that the Comptonizing region structure is entirely different for type-C and type-B QPOs since for type-B QPOs the CompTT model fits for both PCA and PCA+HEXTE spectra resulting in non-physical values. Based on our study, we advocate for the presence of a jet-like structure (a vertically extended Comptonization region) instead of a quasi spherical Compton cloud during the modulation of type-B QPOs.

\subsection{Comparing type-C \& type-B QPO features}

The quasi-simultaneous study suggests that type-C QPOs (e.g., 80146-01-31-00) are associated with strong radio jets ($\sim$34.59 mJy at around $\sim$1.03 hr away), while type-B QPOs (e.g., 80146-01-51-00) are associated with weak radio jets ($\sim$5.59 mJy at around $\sim$2.29hr away). The spectral parameters of observations exhibiting type C/B QPOs suggest that the inner disk radius does not affect the nature of QPOs or that their association is much more complex. Constraining the exact value of R$_{in}$ is difficult as it depends on the chosen model. We are relying mostly on model 3 to compare the nature of type-C/B QPOs, (figures 4-8). The flux ratio between soft and hard components appears to be the only differentiating
factor between type-C \& type-B QPOs. This ratio is proportionally related to the FWHM of these QPOs. From figure 4 we can understand that during the occurrence of type-C QPOs, the ratio of soft to hard flux is low and associate to the narrow FWHM of QPO along with a high flux difference between disk and corona components (Table 3). While in the case of type-B QPOs,
the ratio of soft to hard flux is found to be dominant and connected to the broad FWHM of QPO with less flux difference between disk--corona components (Table 4). We observed an inversely proportional trend (linear fit gives a slope of --0.50$\pm$0.04 and an $F$-test value of 75.96 with a probability of 0.15$e$-06 compared to a constant fit) between the FWHM and soft to hard flux ratio for type-B QPOs i.e. broader type-B QPOs are more influenced by the hard component. The exact physical mechanism causing the broadness of QPO is not known but if we assume that type-B QPOs are produced in the jet, then we speculate that a differential precession of the jet can explain the broadness of the QPO. Though a linear trend is observed (Figure 2), possibly displaying an association between the broadness of type-B QPO and radio flux density, more studies are needed to confirm the scenario of differential precession of the jet.
 Such a differential precession has been used to explain type-C QPOs and their energy dependencies (van den Eijnden et al. 2016, Ingram \& Motta 2019). We noticed that in 80146-01-11-00 and 80146-01-12-00, the inner disk temperature remains the same for different R$_{in}$ values along with the flux values. Hence, irrespective of inner disk radius location, inner disk temperature depends on the disk flux component. During the presence of type-C QPOs, kT$_{in}$ is relatively low compared to type-B QPOs. We find a proportional trend between the inner disk temperature and type-C QPO width (linear fit gives a slope of 0.21$\pm$0.04 and an $F$-test value of 9.06 with a probability of 0.24$e$-01 compared to a constant fit), while for type-B QPOs (linear fit with a slope of 0.05$\pm$0.01 and an $F$-test value of 4.25 with a probability of 0.62$e$-01 compared to a constant fit) (Figure 5). During the occurrence of type-C QPOs, a high flux difference is noted between the disk and corona components, with the corona being dominant. This could be a factor for the relatively
high energetic radio observations that are possibly indicative of a steady jet structure. The opposite scenario is noted in the case
of type-B QPOs which indicates a weak relativistic radio jet (Table 2). The ratio of soft to hard fluxes is found to be increasing with a decrease in radio flux density in the case of type-B QPOs, where the linear fit gives a slope of 13.51$\pm$3.64 and an $F$-test value of 22.49 and a probability of 0.38$e$-03 compared to a constant fit (Figure 6). We also noticed that the values of inner disk temperature are proportionally consistent with the disk flux component as reported previously by McClintock et al. (2009) and Chen et al. (2010). The Comptonization properties are similar in the case of type-B and C QPOs (based on $\Gamma$), which is reasonable as the selected sample has a similar frequency range (Tables 3 and 4).  

Figure 7 shows that soft X-ray flux ($\le$ 4 $\times$ 10$^{-9}$ erg cm$^{-2}$ s$^{-1}$) in type-C QPOs is inversely proportional to the radio flux density. This can be understood in the framework of a truncated accretion disk scenario (Done et al. 2007) i.e., as the disk is moving towards the BH the increase in soft photons cool the corona at the base of a jet. But we did not find any significant change in the inner disk radius, R$_{in}$. Whereas, in type-B QPOs, a dispersed distribution of soft X-ray flux can be seen at low radio flux densities. Figure 8 shows the trend of hard flux vs radio flux density. While in the case of type-C QPOs, an inverse proportionality is noticed,
type-B QPOs show a proportional trend. A linear fit results a slope of 2.06$\pm$0.35 and {an $F$-test value of 43.67 with a probability of 0.17$e$-04 compared to a constant fit}. The trend displayed by type-B QPOs suggests that harder fluxes are connected to higher radio flux densities. This can be explained if the disk is moving away from the BH source or if the base of the jet is increasing in its size. A vertically extended corona can possibly explain both the harder flux values as well as the higher radio emission. We favour the latter scenario since R$_{in}$ was not varying systematically (Table 4). However, more studies are needed to fully understand the suggested scenario.

\subsection{A case study of type-C to type B QPO gradual transition}

The series of QPOs from 80146-01-29-00 to 80146-01-46-00 are falling in the type-C category, some of them showing a relatively higher rms$\%$. ObsIDs 80146-01-48-00 to 80146-01-68-00 are falling in the type-B category with some observations showing no-QPOs in between (except 80146-01-50-00, rapid transition of type-C QPO/No-QPO; Sriram et al. 2021). The ObsID 80146-01-47-00 is a bridge between type-C to type-B QPOs that is seen as type-B QPO in the next ObsID 80146-01-48-00. These are separated by time duration of $\sim$ 24 hr. This transition is marked in the ASM light curve with a star marker and the immediate filled circle shows the type-B QPO occurrence (Fig. 1). This kind of transition can help in understanding the dynamical variations of QPOs and their association with the geometrical structures in the inner region of the accretion disk. We performed spectral (PCA, 3--25 keV) and PDS (6.12--12.69 keV) studies of a and b sections of the first segment, second segment c from 80146-01-47-00, and d section of ObsID 80146-01-48-00 (see Figure 9 for more details). We also studied the PCA+HEXTE (3-100 keV) spectra to understand the contribution from the higher energy band. From the timing analysis, we noticed that in $'bridge  ObsID'$ 80146-01-47-00 a type-C QPO is present at 6.47 Hz with rms $\sim$11.64\% in the beginning. After a short span of $\sim$30 min, QPO frequency has shifted to 7.06 Hz with rms $\sim$9.88\%.
 In a gap of $\sim$ 24 min, the frequency is further slightly shifted to 7.35 Hz, and transformed into a type-B QPO in the next observation 80146-01-48-00 with the frequency of 5.45 Hz, rms $\sim$8.25\%, and a low Q (Table 6). A clear transition of QPO centroid frequency was seen in the dynamical PDS (Figure 10). We show the locations of studied ObsIDs in the hardness intensity diagram (HID), where the hardness ratio is between the 3--6 keV and 7--16 keV energy bands and intensity is in the 3--30 keV energy range. The circle marker represents the type-C, and the square markers represent type-B QPOs along with  dot markers that are used to show the nearest radio observations (see figure 11). The star marker shows the bridge ObsID where QPO transition has been reported (figure 10). Type-B QPOs tend to occupy a softer region in the HID diagram.

It is relatively clear that the production mechanism of type-C and 
type-B QPOs in the accretion disk is entirely different.
Both these types do have a geometric origin, the LT precession model suggests that type-C QPO is due to the precession of hot inner flow (Ingram et al. 2009; Ingram \& Motta 2019),
whereas type-B QPO is possibly due to the jet precession (Motta et al. 2015; 
Stevens \& Uttley 2016; Sriram et al. 2016; de Ruiter 2019) or an extended corona (Wang et al. 2021; Kara et al. 2019).
We explain the gradual variation of type-C to type-B QPO transition, using the 
geometrical model (Table 6). 
Stevens \& Uttley (2016) proposed a precessing jet-like corona that
illuminates the accretion disk to produce type-B QPOs. From Table 6,
sections (a), (b) \& (c) change in most of the parameters are not prominent, and are within error bars. We
can notice that kT$_{in}$ is consistent in these three sections and disk
flux is increasing gradually while hard flux is almost consistent. We notice that during the spectral transition from c to d there is a significant 
increase in both soft flux and inner disk temperature parameters which would cool 
the corona, probably at the base of a jet which is further 
evident from the slight decrease in the hard flux. The weakening of corona is also 
noted from $\Gamma_{simpl}$ and $\Gamma$ using models 2 and 3 from 
sections c to d. Motta et al. (2009) has reported hard to soft state transition
with an increment in $\Gamma_{cutpl}$ and consistent kT$_{in}$ for GX 339-4. Although it was found that R$_{in}$ has slightly increased in models 1 and 3 and decreased in model 2 (Table 6). We conclude that the disk has traversed toward 
the BH since kT$_{in}$ and soft flux have increased and causes the hard flux to decrease.    

Spectral changes suggest that the disk is moving closer to the BH. The LT precession model predicts that the QPO frequency should increase  in this case (Ingram et al. 2009), but the QPO frequency is found to be decreasing, and exhibiting characteristics of type-B QPO (type-C at 7.35 Hz to a type-B QPO at 5.45 Hz from sections c to d). We suggest that during these sections, as the inner disk front is moving towards the BH the corona size decreased, the precession of a jet-like feature might be producing the type-B QPO. Now the question that arises is about the precession of corona at the base of the jet since the coronal parameters have not changed much if we assume model 1 to be the true geometry of the inner accretion disk region. We argue that the corona/hot flow, and the jet are differentially precessing with a slightly different frequency. Such a scenario has been observed in General Relativistic magnetohydrodynamics (GRMHD) simulations (Liska et al. 2019a) where the disk precesses along with the corona as well as with a jet. The precession of the disk is faster compared to that of the corona while the slowest precession is exhibited by the jet component. Hence, the observed type-B QPO could be a combination of the type-C precession in the centroid region of the QPO and the wings of the QPO could be due to the jet precession.

 The other scenario could be that for type-B QPOs, the corona and the disk portions do not maintain the structural integrity to precess due to the tearing of the disk. Such a model is confirmed by smoothed particle hydrodynamics (SPH) (Nixon et al. 2012) and GRMHD simulation (Liska et al. 2019b). Such a portion of the inner region is a common scenario in SIMS. If that is the case, then type-B QPOs should always be accompanied by type-C QPOs which is not seen in any of the BH sources.
Moreover, during the occurrence of type-C QPO, though a steady non-relativistic jet is present, it may not be physically connected to the disk whose precession is the strongest (Liska et al. 2019a). As the disk approaches close to the BH, it connects to the base of the jet (corona) which extends vertically. Modulation in the jet precession is seen as a type-B QPO. There is a redistribution of disk precession and emission powers, to Comptonization region/jet. The optically thick vertical structure of corona is required to explain type-B QPOs (Wang et al. 2021, Stevens \& Uttley 2016). 
From sections c to d, probably the increase in mass accretion rate causes the disk flux to increase but it supports the precession of the Comptonizing region/jet instead of the disk. Such a scenario is observed in the case of MAXI J1348-630 (Zhang et al. 2021).

Based on model 2, the PCA+HEXTE spectral fit results indicate that during type-B QPO (d section), the Comptonization region was harder, compared to type-C QPO (c section; Table 7), but $\Gamma_{simpl}$ = 2.77$\pm$0.06 was softer in d section. This suggests that there exists a soft power-law source in which the disc photons are scattered, resulting in a soft $\Gamma_{simpl}$ component. 
Additionally, there exists a hard power-law source $\Gamma$ = 2.64 $\pm$ 0.04 along with an increase in the hard flux, probably at the base of a jet. Perhaps the disc photons are not able to penetrate and cool down the hard power-law component, possibly due to high optical depth, and most of the photons are scattered at a relatively larger height of a jet. More studies are needed to explain the soft power-law component (Shaw et al. 2016). In a truncated accretion disk scenario, as the disk front moves towards the BH, the hard flux should decrease. Here, the disk front is moving in, but the hard flux has increased with the hardening of $\Gamma$. Moreover, with HEXTE spectral fits alone, $\Gamma$ was found to be harder 2.66$\pm$0.44 in section d, indicating that a relatively hot corona is present. We also noticed that for section d model 1 (i.e., CompTT) resulted in non-physical values suggesting that the Comptonization region was not quasi-spherical during the appearance of type-B QPO. Though the study putatively indicates the presence of a vertical stricture of corona, simultaneous X-ray and radio observations during such a transition can help to corroborate the results.

\section{Conclusions}
We performed spectral and timing studies of a BH source H1743-322 for observations associated with type-C and type-B QPOs to constrain the spectral parameters that indicate the production mechanism of type-B QPOs. Quasi-simultaneous radio observations reported by McClintock et al. (2009) were used to understand the geometrical scenario. The main results are mentioned below.

1. Spectral parameters such as total flux and inner disc radius are not the right parameters to associate with the centroid frequency of type-B QPOs. We noticed that the CompTT model cannot explain the high energy spectrum associated with the type-B QPOs in this source. This indicates that
the geometry of the Comptonization during the type-B QPOs is plausibly different from the configuration associated with type-C QPOs.

2. From timing analysis and quasi-simultaneous radio observations, it is clear that type-B QPO width has an association with the radio flux density. This indicates that a weak/relativistic jet is most likely the origin of type-B QPOs in H1743-322. However, more studies are necessary in this direction.

3. Ratio between soft and hard X-ray fluxes is found to be the differentiating factor between type-C and type-B QPOs. 
High radio flux densities can be noticed with a low X-ray flux ratio in type-C QPOs while type-B QPOs 
are connected to low radio flux along with a high X-ray flux ratio (Fig. 6). As the ratio is increasing, the disk and corona precession are modulated toward the jet precession, possibly causing the type-B QPO. This scenario is supported by the fact that a higher kT$_{in}$ is associated with type-B QPO width (Figure 5) and such QPOs are often connected to the relativistic jets (Fender et al. 2009). The transfer of precession power from disk to corona and jet could be related to the type-B QPO modulation.

4. The probable primary difference between type-C and type-B QPOs is that in type-C QPO only the precession of hot flow/Comptonization region/corona is playing a vital role whereas in type-B either corona and jet are differentially precessing or only a weak jet is precessing. It is important to constrain the location of the inner disk radius during the appearance of type-B QPO, though we know that it is slightly truncated. Future observations would help in elucidating this picture.
 
5. Based on the spectral study, a gradual transition from type-C to type-B QPO is noticed with the hardening of the power-law index ($\Gamma$) and softening of the $\Gamma_{simpl}$. The variation in $\Gamma$ and inner disk radius is not in agreement with the truncated accretion disk geometry. The observed hardening of $\Gamma$ can be explained if a jet is present in the inner region of the accretion disk. This indicates that the Comptonization region has changed in its structure.





\section*{Acknowledgements}
We acknowledge the Referee for providing useful inputs which improved the quality of
the paper. S.H. acknowledges the support from the Senior Research Fellow grant from CSIR-UGC.
K.S. acknowledges the financial support from the SERB Core Research Grant project, Government of India.
This study uses observations made with {\it Rossi X-ray Timing Explorer} (RXTE)
that is obtained through HEASARC Online Service, provided by
NASA/GSFC, in support of the NASA High Energy Astrophysics Programs.

\section*{Data Availability}
Data used in this work can be accessed through the HEASARC website 
(https://heasarc.gsfc.nasa.gov/cgi-bin/W3Browse/w3browse.pl) and is also available with the authors.


\begin{figure}
\begin{minipage}{\columnwidth}
\includegraphics[height=8cm, width=\linewidth, angle=270]{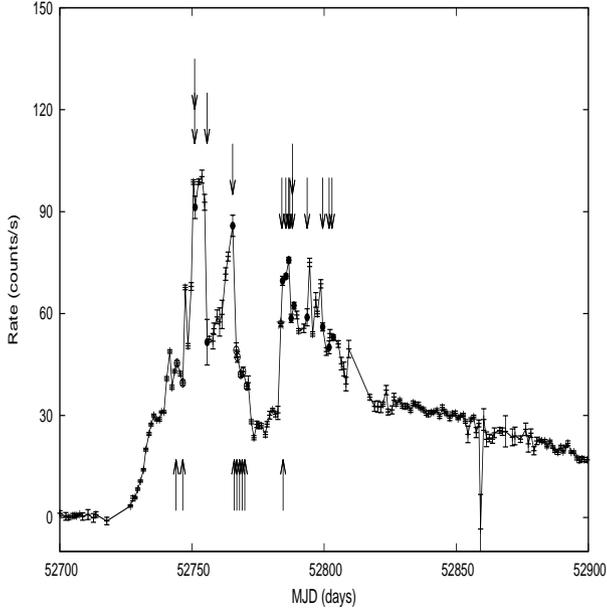} \\
\caption{ASM light curve of 2003 outburst of black hole source H1743-322. Up arrows indicate type-C QPO locations marked with open circles while down arrows indicate type-B QPO locations marked with filled circles. The star symbol indicates the position of ObsID  80146-01-47-00 (for more details see text).}
\end{minipage}
\end{figure}
\begin{figure}
\begin{minipage}{\columnwidth}
\includegraphics[height=8cm, width=\linewidth, angle=270]{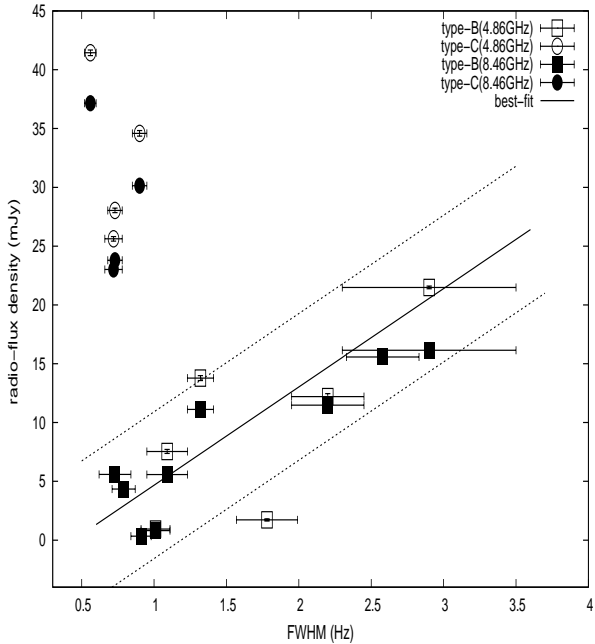} \\
\caption{Distribution of quasi-simultaneous radio flux density (mJy, in 4.86 GHz \& 8.46 GHz) concerning the full width half maximum (FWHM) of type-C (circles) and type-B(boxes) QPOs. Solid line shows the best-fit ($y = (8.35\pm 2.80)*x-(3.68\pm 1.69)$) for type-B QPOs while thin dashed lines show an interval of a 95\% confidence level.}
\end{minipage}
\end{figure}

\begin{figure*}
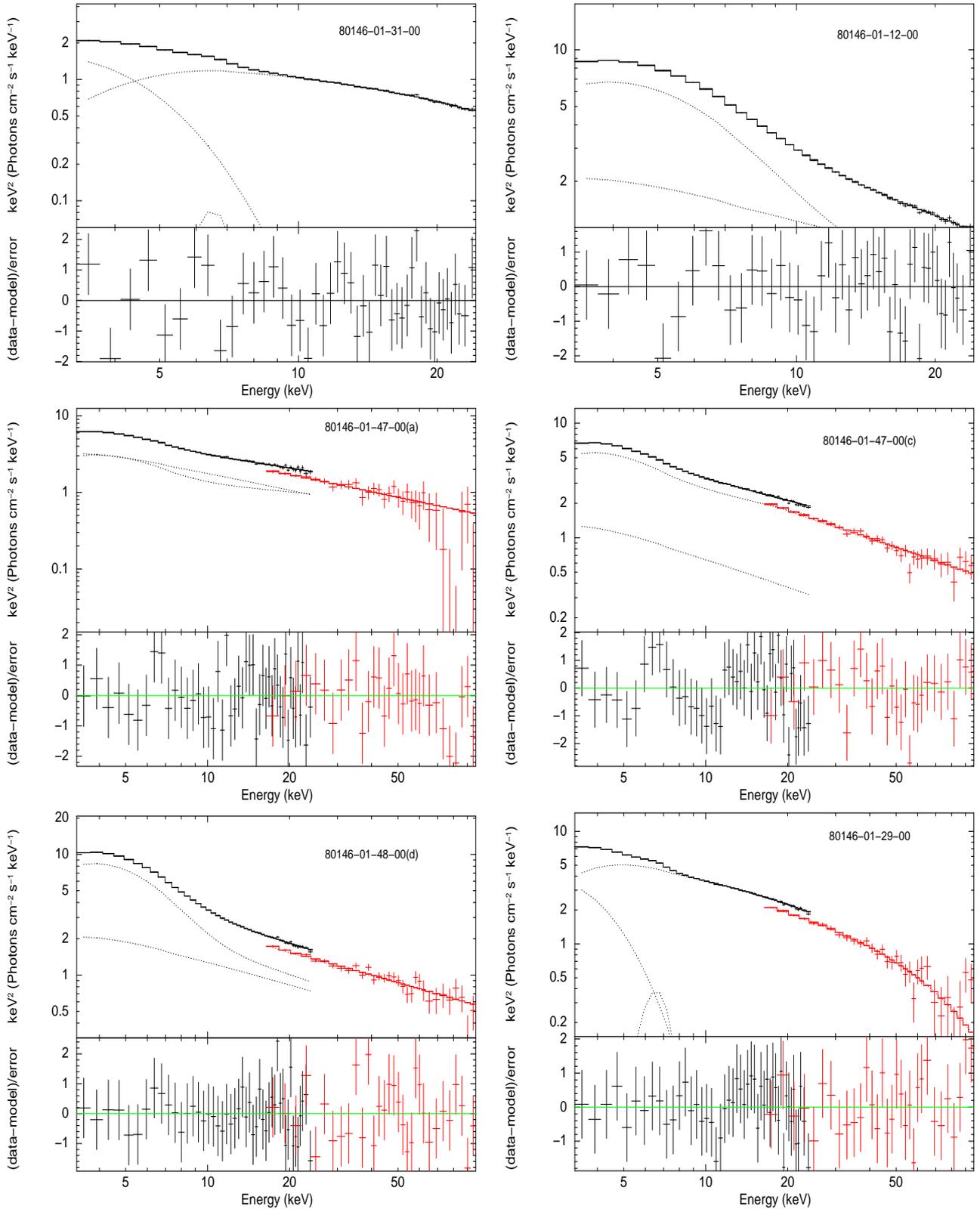

\begin{minipage}{\columnwidth}
\includegraphics[height=8.6cm,width=7cm,angle=-90]{spec_3100_Rv3.ps} 
\end{minipage}
\begin{minipage}{\columnwidth}
\includegraphics[height=8.6cm,width=7cm,angle=-90]{spec_1200_Rv3.ps} 
\end{minipage}

\begin{minipage}{\columnwidth}
\includegraphics[height=8.6cm,width=7cm,angle=-90]{4700A_begin_pcahexte_Rv3.ps} 
\end{minipage}
\begin{minipage}{\columnwidth}
\includegraphics[height=8.6cm,width=7cm,angle=-90]{4700B_pcahexte_Rv3.ps} 
\end{minipage}

\begin{minipage}{\columnwidth}
\includegraphics[height=8.6cm,width=7cm,angle=-90]{4800_pcahexte_Rv3.ps} 
\end{minipage}
\begin{minipage}{\columnwidth}
\includegraphics[height=8.6cm,width=7cm,angle=-90]{spec_2900pcahexte_Rv3.ps} %
\end{minipage}
\caption{ Top panels: Unfolded (PCA) spectra of type-C \& type-B QPOs using model 1 \& 2 respectively. Middle panels: Unfolded (PCA+HEXTE) spectra of a and c sections of bridge ObsID using model 2. Bottom panel: Unfolded (PCA+HEXTE) spectra of d section (type-B QPO, left) using model 2, and 80146-01-29-00(type-C QPO, right) using model 1. The thick lines show the best fits to the spectra and model components are shown with dashed lines.}
       \label{}
\end{figure*}

\begin{figure}
\includegraphics[height=8cm, width=4.5cm, angle=270]{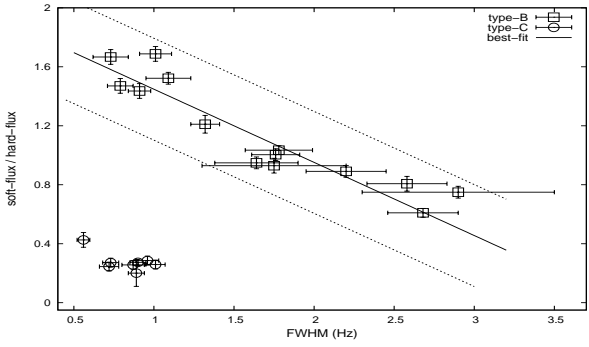} \\
\caption{Width (FWHM) of type-C \& B QPOs vs ratio of soft to hard X-ray fluxes (from model 3). Solid line shows the best-fit for type-B QPOs while thin dashed lines show an interval of a 95\% confidence level.}
\end{figure}
\begin{figure}
\includegraphics[height=8cm, width=4.5cm, angle=270]{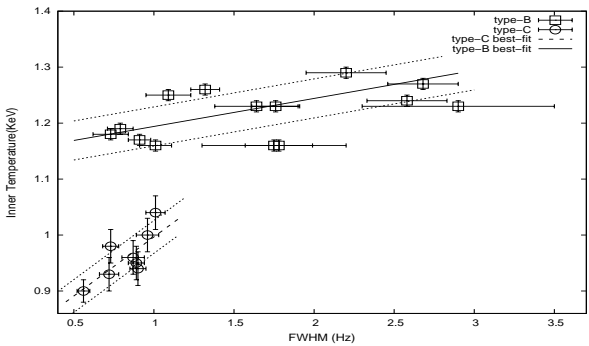} \\ 
\caption{Variation of inner disk temperature (from model 3) concerning FWHM of type-C \& B QPOs. Solid line shows the best-fit for type-B QPOs while thick dashed line shows the best-fit for type-C QPOs and thin dashed lines show  an interval of a 95\% confidence level.}  
\end{figure}
\begin{figure}
\includegraphics[height=8cm, width=4.5cm, angle=270]{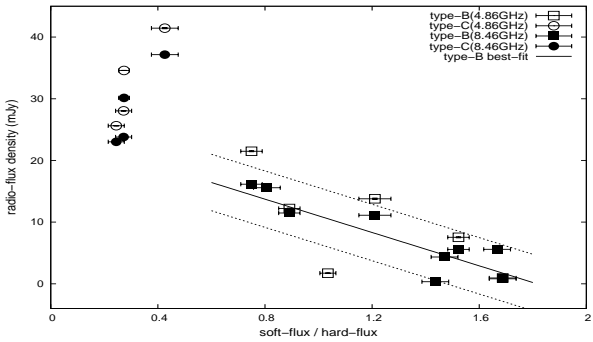} \\
\caption{Quasi-simultaneous radio flux density (mJy, in 4.86 GHz \& 8.46 GHz) vs soft to hard flux ratios (model 3) of type-C \& B QPOs. Solid line shows the best-fit for type-B QPOs while thin dashed lines show an interval of a 95\% confidence level.} 
\end{figure}
\begin{figure}
\includegraphics[height=8cm, width=5cm, angle=270]{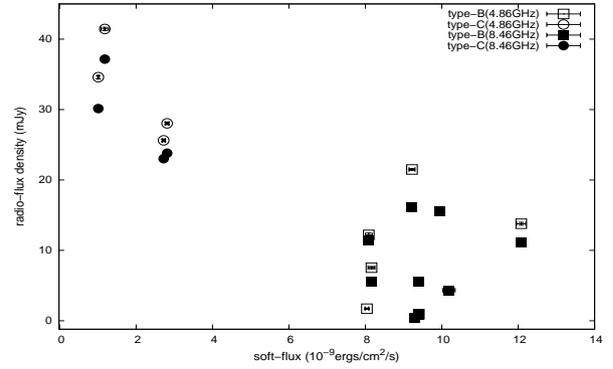} \\ 
\caption{Soft X-ray flux (model 3) vs quasi-simultaneous radio flux density (mJy, in 4.86 GHz \& 8.46 GHz)  }
\end{figure}
\begin{figure}
\includegraphics[height=10cm, width=5cm, angle=270]{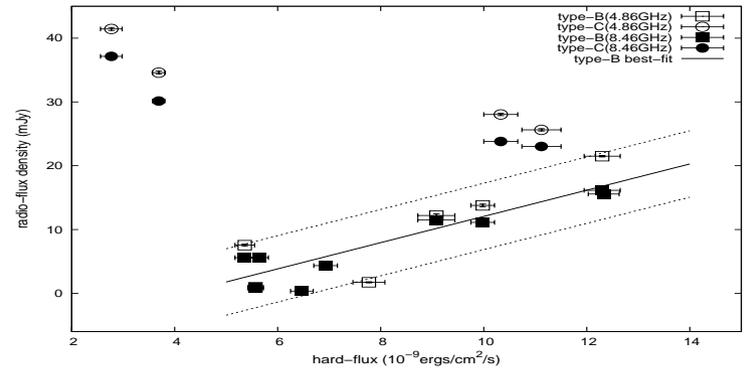} \\ 
\caption{Distribution of hard flux(model 3) with quasi-simultaneous radio flux density (mJy, in 4.86 GHz \& 8.46 GHz). Solid line shows the best-fit for type-B QPOs while thin dashed lines show an interval of a 95\% confidence level.} 
\end{figure}

\begin{figure*}
\centering
\includegraphics[height=14cm, width=7cm, angle=270]{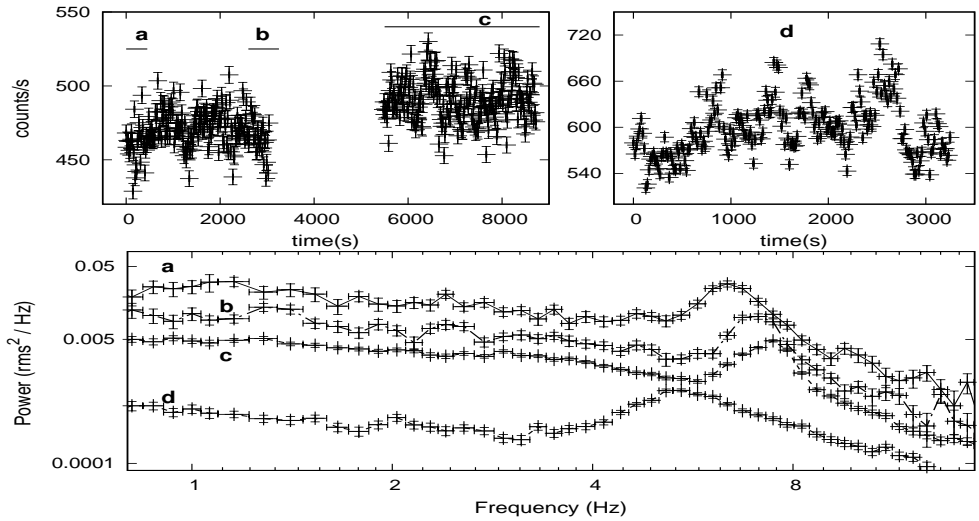}\\ 

\caption{ The top left panel shows the light curve of ObsID 80146-01-47-00 and the top right panel shows the light curve of ObsID 80146-01-48-00 in 6.12--12.69 $keV$ band and sections used for PDS are highlighted with horizontal bars. The bottom panel shows the PDS of the four segments mentioned above and PDS were shifted along Y-axis for clarity. }
\end{figure*}
\begin{figure*}
\centering
\includegraphics[width=0.65\textwidth, angle=270]{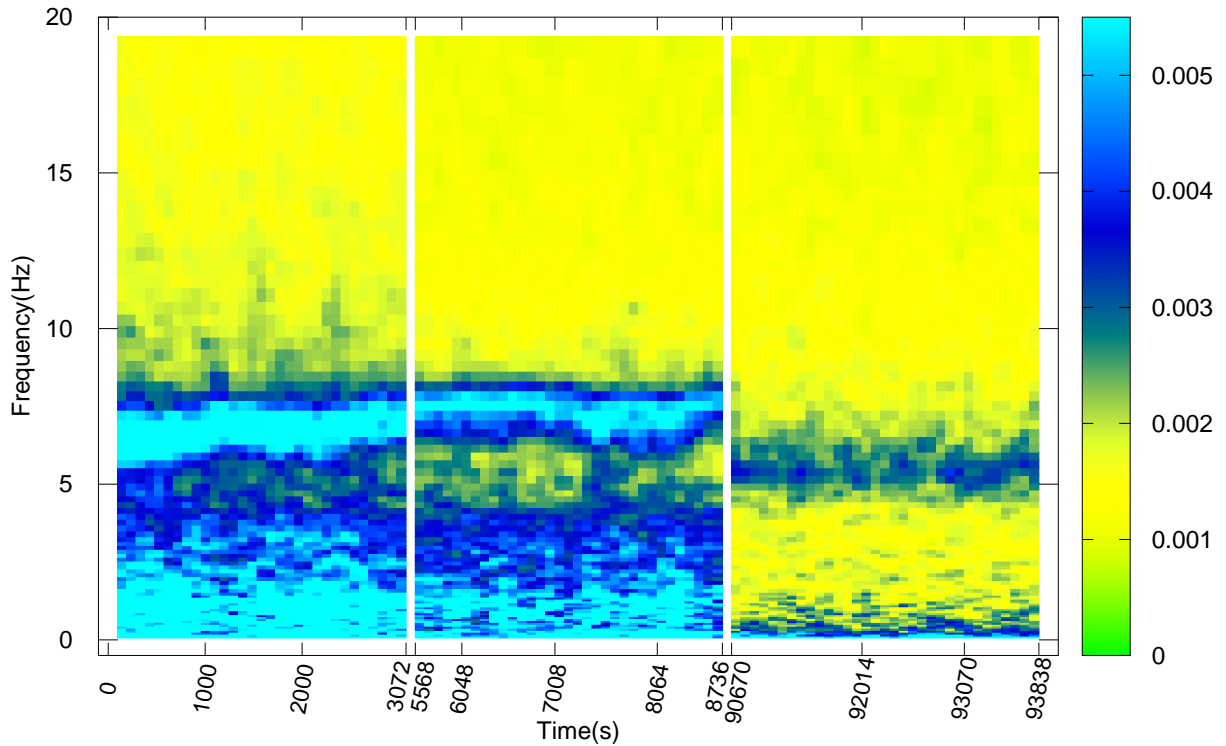}\\ 

\caption{ Dynamic PDS of ObsIDs 80146-01-47-00 and 80146-01-48-00 in 6.12--12.69 keV for the light curve shown in Fig. 9. Time bin is of 96 s duration along with a frequency bin of 1/512 Hz. A strong type-C QPO at 6.47 Hz was observed during the initial variation and finally settled as  type-B QPO at 5.45 Hz. The colour scale depicts the strength of the power of the QPO.  }
\end{figure*}

\clearpage
\begin{figure*}
\centering
\includegraphics[height=14cm, width=10cm, angle=270]{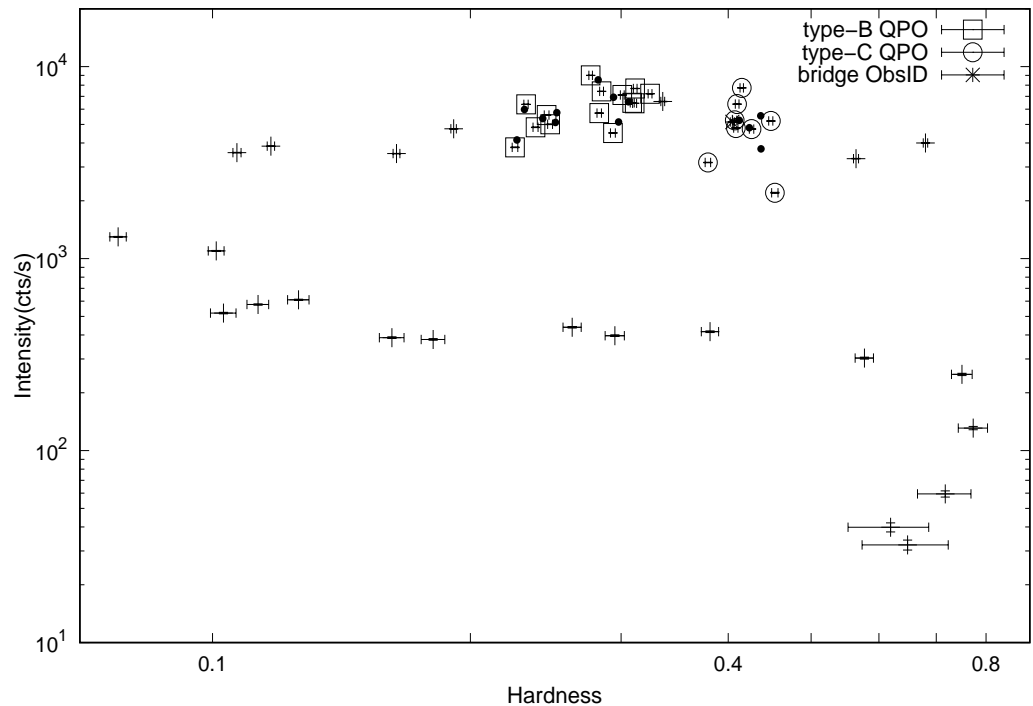}\\ 

\caption{ HID for the 2003 outburst of H1743-322. Hardness is defined as 7--16/3--6 keV. Intensity
is given in the 3--30 keV energy range. Boxes mark type-B QPOs, circles mark type-C QPOs, star marks bridge ObsID and black dots mark radio events.}
\end{figure*}

\clearpage

\begin{table}
\begin{minipage}[t]{\columnwidth}
\scriptsize
\caption{Log of the QPO parameters associated with the Lorentzian fits. PDS power unit is (rms/mean)$^{2}$ Hz$^{-1}$. The harmonic QPO parameters were reported in the parenthesis.}
\centering
\begin{tabular}{cc|c|c|c|c|c}
\hline
\hline
    
 Obs ID & Energy band ($keV$) & LC\footnote{Centroid frequency of the Lorentzian component} & LW\footnote{Width of the Lorentzian component} & Q\footnote{$\frac{\nu}{\delta \nu}$, Quality factor of fundamental frequency} & rms(\%)\footnote{rms amplitude} & red. $\chi^{2}$(dof)\footnote{Degree of freedom}\\
\hline
\hline
 & & &  Type-C QPOs & &  &\\
\hline 	
\hline									
80146-01-02-00 & 3.68-5.71 & 5.58$\pm$0.02\footnote{fundamental centroid frequency} &  0.77$\pm$0.06  &  7.25$\pm$0.57   & 7.47$\pm$0.19  & 1.50(130)\\
   & 6.12-12.69 & 5.64$\pm$0.02 & 0.89$\pm$0.05  &  6.31$\pm$0.36 & 12.23$\pm$0.42  & 1.39(130) \\	
\hline							
80146-01-29-00 & 3.68-5.71& 5.54$\pm$0.02 &  0.57$\pm$0.05  &  9.71$\pm$0.85   & 6.45$\pm$0.31  & 1.35(130) \\
   & 6.12-12.69 & 5.60$\pm$0.02 & 0.87$\pm$0.07 &  6.43$\pm$0.52 & 11.53$\pm$0.72  & 1.33(130)\\
\hline
80146-01-30-00 & 3.68-5.71 & 4.44$\pm$0.05(8.79$\pm$0.15)\footnote{harmonic centroid frequency}  &  0.69$\pm$0.05(2.44$\pm$0.44)  &  6.46$\pm$0.47(3.60$\pm$0.65)   & 9.35$\pm$0.23(4.07$\pm$0.98)  & 1.20(130) \\
   & 6.12-12.69 & 4.44$\pm$0.07(8.55$\pm$0.15) & 0.72$\pm$0.06(2.14$\pm$0.40)  &  6.17$\pm$0.52(3.99$\pm$0.75) & 15.05$\pm$0.41(5.80$\pm$0.70)  & 1.14(130)\\ 
\hline
80146-01-31-00 & 3.68-5.71 & 5.37$\pm$0.02 & 0.74$\pm$0.05  &  7.26$\pm$0.49  & 8.12$\pm$0.18  & 1.34(130) \\
   & 6.12-12.69 & 5.41$\pm$0.02 & 0.90$\pm$0.05  &  6.01$\pm$0.33 & 14.21$\pm$0.82  & 1.55(130)\\
\hline
80146-01-32-00 & 3.68-5.71 & 4.88$\pm$0.01  & 0.58$\pm$0.04   & 8.17$\pm$0.58   & 8.82$\pm$0.52  & 1.50(130) \\
   & 6.12-12.69 & 4.81$\pm$0.12(9.43$\pm$0.13) & 0.73$\pm$0.05(1.63$\pm$0.39)  &  6.56$\pm$0.48(5.78$\pm$1.38) & 14.60$\pm$0.48(5.58$\pm$0.42)  & 1.18(130)\\
\hline
80146-01-33-01 & 3.68-5.71 & 5.94$\pm$0.03  & 0.81$\pm$0.17  & 7.32$\pm$1.53   & 6.33$\pm$0.16  & 1.04(130) \\
   & 6.12-12.69 & 6.05$\pm$0.03 & 0.96$\pm$0.07  &  6.30$\pm$0.46 & 13.11$\pm$0.22 & 1.29(42) \\
\hline
80146-01-33-00 & 3.68-5.71 & 6.17$\pm$0.03  &  0.89$\pm$0.08   &  6.96$\pm$0.62   & 6.39$\pm$0.28  & 0.87(130) \\
   & 6.12-12.69 & 6.26$\pm$0.02 & 1.01$\pm$0.06  &  6.23$\pm$0.37 & 12.38$\pm$0.18 & 1.29(130) \\
\hline
80146-01-03-00 & 3.68-5.71 & 4.72$\pm$0.02  &  0.60$\pm$0.04   &  7.90$\pm$0.53   & 8.92$\pm$0.79  & 1.02(163) \\
   & 6.12-12.69 & 4.75$\pm$0.02 & 0.56$\pm$0.04  & 8.49$\pm$0.61 & 13.34$\pm$0.35 & 1.66(163) \\
\hline
80146-01-47-00 & 3.68-5.71 & 7.01$\pm$0.09  & 1.01$\pm$0.19   &  6.94$\pm$1.30   & 4.79$\pm$0.31  & 1.20(163) \\
   & 6.12-12.69 &  $7.06\pm$0.06 &  1.03$\pm$0.12  &  6.85$\pm$0.80 & 9.88$\pm$0.12 & 1.24(130) \\
\hline
\hline
 & & &  Type-B QPOs & & & \\
\hline 	
\hline								
80146-01-11-00 & 3.68-5.71 & 5.49$\pm$0.08   &  2.66$\pm$0.24  &  2.06$\pm$0.19   & 4.95$\pm$0.25 & 1.08(130) \\
   & 6.12-12.69 & 5.60$\pm$0.06 & 2.68$\pm$0.22  & 2.10$\pm$0.17 & 8.75$\pm$0.25  & 0.94(130)  \\
\hline
80146-01-12-00 & 3.68-5.71 & 5.77$\pm$0.09  &  2.66$\pm$0.29  &  2.17$\pm$0.24   & 3.77$\pm$0.21   & 1.18(130)  \\
   & 6.12-12.69 & 5.87$\pm$0.07 & 2.20$\pm$0.25 &  2.67$\pm$0.30 & 8.08$\pm$0.27   & 1.30(130)  \\
\hline
80146-01-48-00 & 3.68-5.71 & 5.37$\pm$0.09(8.12$\pm$0.33)  &  2.24$\pm$0.32(1.91$\pm$0.76)  &  2.40$\pm$0.34(4.25$\pm$1.66)   & 4.41$\pm$0.17(2.09$\pm$0.11)  & 0.96(130) \\
   & 6.12-12.69 & 5.45$\pm$0.15 & 1.64$\pm$0.26  &  3.55$\pm$0.53 & 8.25$\pm$0.13   & 1.31(130) \\
\hline
80135-02-02-00 & 3.68-5.71 & 5.54$\pm$0.07  & 1.99$\pm$0.21  &  2.79$\pm$0.29  & 3.81$\pm$0.19  & 0.96(130)\\
   & 6.12-12.69 & 5.54$\pm$0.05 & 1.76$\pm$0.15  &  3.14$\pm$0.27 & 8.19$\pm$0.23  & 1.39(130) \\
\hline
80144-01-01-02 & 3.68-5.71 & 5.39$\pm$0.12  &  1.85$\pm$0.39 &  2.91$\pm$0.62   & 4.15$\pm$0.24  & 0.78(130) \\
   & 6.12-12.69 & 5.51$\pm$0.09 & 1.75$\pm$0.45  &  3.15$\pm$0.81 & 8.03$\pm$0.18  & 0.80(130)\\
\hline
80146-01-68-00 & 3.68-5.71 & 5.31$\pm$0.09  &  2.22$\pm$0.40  &  2.39$\pm$0.43   & 3.80$\pm$0.22  & 0.81(130) \\
   & 6.12-12.69 & 5.35$\pm$0.07 & 1.78$\pm$0.21  &  3.01$\pm$0.35 & 6.23$\pm$0.17  & 1.37(130) \\
 \hline
80144-01-01-00 & 3.68-5.71 & 5.33$\pm$0.14  &  3.03$\pm$0.44  &  1.76$\pm$0.26   & 3.94$\pm$0.18  & 1.29(163) \\
   & 6.12-12.69 & 5.32$\pm$0.13 & 2.90$\pm$0.60  &  1.83$\pm$0.38 & 6.91$\pm$0.48  & 0.96(163) \\
 \hline
80146-01-16-00 & 3.68-5.71 & 4.56$\pm$0.20  &  1.04$\pm$0.37  &  4.40$\pm$1.57   & 1.87$\pm$0.23  & 1.47(163) \\
   & 6.12-12.69 & 4.61$\pm$0.04(8.89$\pm$0.16) & 1.09$\pm$0.14(3.32$\pm$0.59)  &  4.21$\pm$0.54(2.68$\pm$0.47) & 4.79$\pm$0.21(4.00$\pm$0.12)  & 1.04(163) \\
\hline
80146-01-27-00 & 3.68-5.71 & 5.84$\pm$0.28  &  3.46$\pm$0.56  & 1.69$\pm$0.23  & 2.01$\pm$0.28  & 1.42(95) \\
   & 6.12-12.69 & 5.29$\pm$0.03 & 1.32$\pm$0.09  &  4.01$\pm$0.27 & 5.65$\pm$0.18  & 1.73(163) \\
\hline
80146-01-49-00 & 3.68-5.71 & 5.12$\pm$0.08  &  2.11$\pm$0.29  &  2.43$\pm$0.33   & 3.87$\pm$0.37  & 1.44(163) \\
   & 6.12-12.69 & 5.37$\pm$0.06 & 2.58$\pm$0.25  &  2.08$\pm$0.20 & 7.07$\pm$0.33  & 1.21(163) \\
\hline
80146-01-51-00 & 3.68-5.71 & 4.49$\pm$0.11(9.05$\pm$0.34)  &  1.14$\pm$0.26(1.49$\pm$0.62)  & 3.95$\pm$0.90(6.06$\pm$2.52)   & 2.23$\pm$0.17(1.81$\pm$0.23)  & 1.55(163) \\
   & 6.12-12.69 & 4.58$\pm$0.05(8.87$\pm$0.09) & 0.73$\pm$0.11(2.02$\pm$0.22)  &  6.29$\pm$0.94(4.40$\pm$0.48) & 4.79$\pm$0.33(4.58$\pm$0.22)  & 1.16(163) \\
\hline
80146-01-56-00 & 3.68-5.71 & 4.82$\pm$0.05(9.62$\pm$0.18)  &  0.75$\pm$0.12(0.91$\pm$0.38)  &  6.45$\pm$1.03(10.58$\pm$4.41)   & 2.49$\pm$0.19(1.48$\pm$0.21)  & 1.34(163) \\
   & 6.12-12.69 & 4.83$\pm$0.03(9.33$\pm$0.13) & 0.79$\pm$0.08(2.21$\pm$0.31)  &  6.15$\pm$0.62(4.22$\pm$0.60) & 5.74$\pm$0.38(4.79$\pm$0.24)  & 1.16(163) \\
\hline
80146-01-62-00 & 3.68-5.71 & 4.46$\pm$0.07(9.10$\pm$0.22)  &  1.21$\pm$0.20(2.19$\pm$0.49)  &  3.67$\pm$0.61(4.16$\pm$0.94)   & 2.30$\pm$0.40(1.67$\pm$0.25)  & 0.94(163) \\
   & 6.12-12.69 & 4.51$\pm$0.04(8.92$\pm$0.10) & 1.01$\pm$0.10(2.69$\pm$0.26)  &  4.47$\pm$0.44(3.32$\pm$0.32) & 4.69$\pm$0.44(4.79$\pm$0.31)  & 1.10(163) \\
\hline
80146-01-65-00 & 3.68-5.71 & 4.94$\pm$0.05  &  1.00$\pm$0.08  & 4.95$\pm$0.11    & 2.64 $\pm$0.40  & 0.81(163) \\
   & 6.12-12.69 & 4.94$\pm$0.03(9.51$\pm$0.11) & 0.91$\pm$0.07(2.50$\pm$0.28)  &  5.45$\pm$0.41(3.80$\pm$0.43) & 5.96$\pm$0.31 (4.51$\pm$0.33) & 1.18(163) \\
\hline
\hline
\end{tabular}
\\
\end{minipage}
\end{table}%

\clearpage
\begin{table}
\begin{minipage}[t]{\columnwidth}
\scriptsize
\caption{Log of X-ray and radio observations for the selected sample. Radio flux densities (R) were taken from McClintock et al (2009).} 
\label{ }
\centering 
\begin{tabular}{ccccccc}
\hline
\hline
ObsID  (QPO type) & X-ray MJD & Radio MJD (nearest) & R (4.86 GHz, mJy) & Radio MJD (nearest) & R (8.46 GHz, mJy)& Time difference (hours)\\
\hline
\hline
80146-01-02-00(C)	& 52744.2206 & 52745.4375 	&	41.44$\pm$0.24 & 52745.4336	 &	37.15$\pm$0.13 & $>$1 day\\
80146-01-29-00(C)	& 52766.5927 & 52767.5195	&	25.62$\pm$0.19 & 52767.5195	&	23.02$\pm$0.12 & 22.24 h \\
80146-01-30-00(C)	& 52767.8657 & 52767.5195 	&	25.62$\pm$0.19 & 52767.5195	 &	23.02$\pm$0.12 & 8.30 h\\
80146-01-31-00(C)	& 52768.5472 & 52768.5039	&	34.59$\pm$0.24 & 52768.4922	&	30.14$\pm$0.16 & 1.03 h/1.32\\
80146-01-32-00(C)	& 52769.7466 & 52769.5156 	&	28.04$\pm$0.19 & 52769.5078	 &	23.80$\pm$0.14 & 5.54 h/5.73\\
80146-01-33-01(C)	& 52770.3880 & 52769.5156	&	28.04$\pm$0.19 & 52769.5078	&	23.80$\pm$0.14 & 20.93 h/21.12 h\\
80146-01-33-00(C)	& 52770.6716 & 52769.5156	&	28.04$\pm$0.19 & 52769.5078	&	23.80$\pm$0.14 & $>1$ day\\
80146-01-03-00(C)	& 52746.1959 & 52745.4375 	&	41.44$\pm$0.24 & 52745.4336	 &	37.15$\pm$0.13 & 18.20 h/18.29 h\\
80146-01-47-00(C)	& 52783.5416 & 52786.3477  	&  21.49$\pm$0.12\&22.49$\pm$0.41  & 52786.0000  &	15.57$\pm$0.30\&15.87$\pm$0.32  & $>1$ day\\
\hline

80146-01-11-00(B)	& 52751.7252 &	52752.5352	& 	12.20$\pm$0.26 &	52752.5352	&	11.48$\pm$0.20 & 19.44 h\\
80146-01-12-00(B)	& 52751.9921 &	52752.5352	&	12.20$\pm$0.26 &	52752.5352	&	11.48$\pm$0.20 & 13.03 h \\
80146-01-48-00(B)	& 52784.5757 &	52786.3477	&	21.49$\pm$0.12 & 52786.0000	&	15.57$\pm$0.30 & $>$1 day\\
80135-02-02-00(B)	& 52787.6521 &	52786.3477	&	21.49$\pm$0.12 & 52788.3750	&	5.59$\pm$0.05 & ($>$1 day)/17.35 h\\
80144-01-01-02(B)	& 52788.0257 &	52786.3477	&	21.49$\pm$0.12 & 52788.3750	&	5.59$\pm$0.05 & ($>$1 day)/8.38 h\\
80146-01-68-00(B)	& 52804.6291 &	52809.2148	&	1.72$\pm$0.09 & 52801.2969	&	0.34$\pm$0.04 & $>$1 day \\
80144-01-01-00(B)	& 52786.8502 &	52786.3477  &  21.49$\pm$0.12\&22.49$\pm$0.41  & 52786.4844  &  16.90$\pm$0.38\&16.15$\pm$0.16  & 12.06/8.78 h \\
80146-01-16-00(B)	& 52755.9602 &	52755.5312  &   7.54$\pm$0.17 		       & 52755.4961  &	5.58$\pm$0.12 			& 10.30/11.14 h \\
80146-01-27-00(B)	& 52764.9070 &	52765.4375  &	13.78$\pm$0.21		       & 52765.4219  &	11.12$\pm$0.13			& 12.73/12.36 h \\
80146-01-49-00(B)	& 52785.4831 &	52786.3477  &  21.49$\pm$0.12\&22.49$\pm$0.41  & 52786.0000  &	15.57$\pm$0.30\&15.87$\pm$0.32  & 20.75/12.41 h \\
80146-01-51-00(B)	& 52788.4706 &	--  	    &	    --	  		       & 52788.3750  &	5.59$\pm$0.05 			& 2.29 h \\
80146-01-56-00(B)	& 52793.6080 &	--  	    &	    --	  		       & 52794.4023  &	4.34$\pm$0.04		        & 19.06 h \\
80146-01-62-00(B)	& 52799.4708 &	52800.2383  &	0.94$\pm$0.09  		       & 52799.2539  &	0.81$\pm$0.06		        & 18.42/5.21 h \\
80146-01-65-00(B)	& 52801.9153 &	52800.2383  &	0.94$\pm$0.09 		       & 52801.2969  &	0.34$\pm$0.04		        & ($>$1 day)/14.84 h \\
\hline
\end{tabular}
\end{minipage}
\end{table}

\clearpage
\begin{landscape}
\begin{table}
\begin{minipage}[t]{\columnwidth}
\caption{Spectral parameters of different observations exhibiting type-C QPOs. Fluxes are in units 10$^{-9}$ erg cm$^{-2}$ s$^{-1}$. } 
\label{ }
\tiny
\centering
\begin{tabular}{ccccccccc}
\hline
\hline
sections & 80146-01-02-00 & 80146-01-29-00 & 80146-01-30-00 & 80146-01-31-00 & 80146-01-32-00 & 80146-01-33-01 & 80146-01-33-00 & 80146-01-03-00\\
\hline

 & & & & model 1 & &  & \\

\hline
$kT_{in}$ ($keV$)\footnote{inner disk temperature}	& 0.82$\pm$0.02 & 0.98$\pm$0.03	& 0.92$\pm$0.03	        & 0.84$\pm$0.02	& 1.01$\pm$0.03 &  1.03$\pm$0.03	& 1.07$\pm$0.04 & 0.81$\pm$0.02\\
 
N$_{diskbb}$	& 731$\pm$81 & 632$\pm$112	& 636$\pm$115	& 753$\pm$83 & 424$\pm$70 &  508$\pm$91	& 404$\pm$77 & 704$\pm$87 \\

kT$_{0}$\footnote{Input soft photon temperature} & 1.04 &  0.43 & 0.49	& 1.05  & 0.34 & 0.31   & 0.41  & 0.97\\
kT$_{e}$\footnote{electron temperature}($keV$) & 8.33$\pm$1.05 &  9.15$\pm$1.65 &	7.63$\pm$0.83   & 7.86$\pm$1.04  & 7.88$\pm$0.93 & 8.34$\pm$1.53  &  10.19$\pm$2.78 & 5.87$\pm$0.46\\
$\tau$\footnote{optical depth}	& 5.10$\pm$0.53 &  4.86$\pm$0.72 & 5.84$\pm$0.55 & 5.63$\pm$0.62& 5.76$\pm$0.57 & 5.33$\pm$0.79 &  4.50$\pm$0.90 & 7.14$\pm$0.53\\
N$_{comptt}$	& 0.06$\pm$0.01 &   0.68$\pm$0.08	& 0.55$\pm$0.03	& 0.07$\pm$0.01	& 0.74$\pm$0.04 & 1.00$\pm$0.09		& 0.62$\pm$0.10 & 0.08$\pm$0.01\\
diskbb flux	& 1.13$\pm$0.09 & 3.04$\pm$0.19	& 2.18$\pm$0.29	& 1.46$\pm$0.09 	& 2.52$\pm$0.17 & 3.26$\pm$0.20	& 3.22$\pm$0.10 & 1.06$\pm$0.09\\
CompTT flux	& 3.00$\pm$0.10 & 13.36$\pm$0.20	& 11.68$\pm$0.11	& 3.25$\pm$0.08	& 10.66$\pm$0.08 & 12.51$\pm$0.13	&  13.15$\pm$0.08 & 2.66$\pm$0.04\\
$\chi^{2}$ / dof  & 33 / 41 &  29 / 41		& 32/41	&  43 / 41	& 26 / 41 & 34 / 41& 43 / 41 &50/41\\
R$_{in}$(km)& 54$\pm$3 & 50$\pm$4& 51$\pm$5& 55$\pm$3 & 41$\pm$3 & 45$\pm$4	& 40$\pm$4 & 53$\pm$3\\
\hline
 & & & & model 1, kT$_{0}=$$kT_{in}$ & &  & \\
\hline
$kT_{in}$ ($keV$) & 0.73$\pm$0.02 	&  0.77$\pm$0.02	& 0.76$\pm$0.02	&  0.78$\pm$0.02	& 0.81$\pm$0.03 	& 0.82$\pm$0.03	& 0.80$\pm$0.02 &0.77$\pm$0.02 \\
N$_{diskbb}$	 & 857$\pm$120 &  2867$\pm$375 & 2337$\pm$326	&  830$\pm$108	& 1806$\pm$234 & 2236$\pm$309	& 2334$\pm$320 & 707$\pm$92 \\
kT$_{0}$   	& $=kT_{in}$ 		&  $=kT_{in}$	& $=kT_{in}$	& $=kT_{in}$	& $=kT_{in}$ 	& $=kT_{in}$	& $=kT_{in}$ & $=kT_{in}$ \\
kT$_{e}$   	& 10.36$\pm$2.75 	&  10.15$\pm$2.62	& 7.86$\pm$0.90	&  8.98$\pm$1.91	& 8.50$\pm$1.30 	& 9.58$\pm$2.88	& 14.18$\pm$6.50 &6.15$\pm$0.60 \\
$\tau$   	 & 4.14$\pm$0.87 	&  4.41$\pm$0.85	& 5.68$\pm$0.56	&  4.95$\pm$0.76	& 5.35$\pm$0.68 	& 4.68$\pm$1.05	& 3.30$\pm$1.13  & 6.75$\pm$0.61\\
N$_{comptt}$  	& 0.08$\pm$0.01 	&  0.28$\pm$0.06	& 0.32$\pm$0.02	&  0.01$\pm$0.01	& 0.25$\pm$0.03 	& 0.27$\pm$0.05	& 0.20$\pm$0.08 & 0.10$\pm$0.01\\
$\chi^{2}$ / dof& 36 / 41  		&  32 / 41	& 33/41		&  43 / 41	& 32 / 41 	& 35 / 41	& 45 / 41 & 55 / 41\\
diskbb flux     & 0.67$\pm$0.08 		&  3.35$\pm$0.12		& 2.48$\pm$0.08	&  1.01$\pm$0.07		& 2.68$\pm$0.19 		& 3.43$\pm$0.10		& 3.35$\pm$0.21 & 0.82$\pm$0.09\\
CompTT flux    & 3.44$\pm$0.11 		&  13.00$\pm$0.12	& 11.35$\pm$0.20		&  3.68$\pm$0.21		& 10.43$\pm$0.31 	& 12.25$\pm$0.12		& 12.86$\pm$0.09 & 2.88$\pm$0.22\\
R$_{in}$(km) 	& 59$\pm$4 	&  108$\pm$7	& 97$\pm$7	&  57$\pm$4	& 85$\pm$5 	& 95$\pm$7	& 96$\pm$7 & 53$\pm$3\\
\hline
$\nu /Q_{\nu}$  & 5.64$\pm$0.02 / 6.31 &  5.60$\pm$0.02 / 6.43	& 4.44$\pm$0.07 / 6.17	& 5.41$\pm$0.02 / 6.01   & 4.81$\pm$0.12 / 6.56 & 6.05$\pm$0.03 / 6.30  & 6.26$\pm$0.02 / 6.23 & 4.75$\pm$0.02 / 8.48 \\
\hline
 & & & & model 3 & &  & \\
\hline
 $kT_{in} (keV)$ &0.95$\pm$0.03	 & 0.96$\pm$0.03     & 0.93$\pm$0.03    & 0.94$\pm$0.03   & 0.98$\pm$0.03    & 1.00$\pm$0.03     &1.04$\pm$0.03 &0.90$\pm$0.02\\
 N$_{diskbb}$	& 165$\pm$40 & 790$\pm$152 & 766$\pm$127 & 265$\pm$52 & 563$\pm$96 & 633$\pm$103 & 497$\pm$118 & 319$\pm$39 \\
 $\Gamma$\footnote{Power-law index}	& 2.26$\pm$0.10  & 2.17$\pm$0.12     & 1.93$\pm$0.13    & 2.10$\pm$0.15   & 1.97$\pm$0.13    & 2.09$\pm$0.16     &2.23$\pm$0.13	& 1.71$\pm$0.18\\
 $E_{cut}$\footnote{high-energy cutoff} $(keV)$& 28.10$\pm$6.27 & 28.03$\pm$7.12    & 21.07$\pm$4.11   & 25.29$\pm$6.89  & 23.19$\pm$4.82   & 25.46$\pm$7.23    &33.41$\pm$10.29 &14.39$\pm$2.71	\\
 $N_{cutpl}$	& 2.41$\pm$0.44  & 7.68$\pm$1.67     & 4.46$\pm$1.05    & 1.93$\pm$0.38   & 4.37$\pm$0.99    & 6.37$\pm$1.75     &8.19$\pm$1.79	&0.88$\pm$0.25\\
 $\chi^{2}$ / dof& 35/41	 & 31/41	     & 33/41	        & 43/41	          & 28/41 	     & 34/41		    &44/41 &53/41	\\
 R$_{in}(km)$	& 26$\pm$3 & 56$\pm$5    & 55$\pm$5   & 33$\pm$3  & 48$\pm$4   & 50$\pm$4    &45$\pm$5 &36$\pm$2\\ 
 diskbb flux	& 0.69$\pm$0.33	& 3.34$\pm$0.07	& 2.72$\pm$0.04	& 1.01$\pm$0.02	& 2.81$\pm$0.05	    & 3.50$\pm$0.08 	            &3.36$\pm$0.07 & 1.18$\pm$0.10\\ 
 cutoffpl flux	& 3.44$\pm$0.08	& 13.04$\pm$0.41	& 11.12$\pm$0.38	& 3.69$\pm$0.12	& 10.33$\pm$0.33	     & 12.24$\pm$0.31	 &12.99$\pm$0.36 &2.77$\pm$0.21\\ 
\hline
\end{tabular}
\end{minipage}
\end{table}
\end{landscape}
\clearpage

\clearpage
\begin{table}
\begin{minipage}[t]{\columnwidth}
\scriptsize
\caption{Spectral parameters of observations exhibiting type-B QPOs. Fluxes are in units of 10$^{-9}$ erg cm$^{-2}$ s$^{-1}$.}
\label{ }
\tiny
\centering
\begin{tabular}{ccccccccc}
\hline
\hline
sections & 80146-01-11-00 & 80146-01-12-00   & 80135-02-02-00  & 80144-01-01-02  & 80146-01-68-00 & 80144-01-01-00 & 80146-01-16-00 \\
\hline

 &  &   &   &  model 2  & & & \\ 
\hline

$\Gamma_{simpl}$\footnote{Power-law index of SIMPL model}	 &  3.36$\pm$0.74	& 3.33$\pm$0.47	 & 2.69$\pm$0.61	& 2.48$\pm$0.10	& 2.14$\pm$0.13 & 2.82$\pm$0.68   & 3.59$\pm$0.86 \\
f$_{scat}$\footnote{The scattered fraction of soft seed photons} &   0.18$\pm$0.05	        &  0.20$\pm$0.04		 &  0.10$\pm$0.04  & 0.17$\pm$0.06 & 0.03$\pm$0.01 & 0.13$\pm$0.05   & 0.12$\pm$0.05  \\
$kT_{in}$ ($keV$)							 & 1.19$\pm$0.02 & 1.21$\pm$0.02  & 1.20$\pm$0.02  & 1.15$\pm$0.03 & 1.16$\pm$0.01 & 1.19$\pm$0.02   & 1.20$\pm$0.01 \\
N$_{diskbb}$ &  163$\pm$17 &  582$\pm$42  &  751$\pm$55 &  402$\pm$102 & 601$\pm$47 & 723$\pm$63 & 575$\pm$50 \\
$\Gamma$\footnote{Power-law index}							&   2.80$\pm$0.14   &   2.68$\pm$0.25   &   2.68$\pm$0.17 &  3.42$\pm$0.13 & 2.62$\pm$0.11 & 2.60$\pm$0.17   & 2.57$\pm$0.33    \\
N$_{pl}$   &  4.29$\pm$0.53   &  6.03$\pm$1.12   &  8.44$\pm$2.67  &  16.11$\pm$6.00 & 7.84$\pm$1.93 & 8.29$\pm$2.65   & 2.98$\pm$0.99 \\
$\chi^{2}$ / dof  &  36 / 41   &   37 / 43    &   36 / 43    &   50 / 39 & 32 / 43 & 29/43  	    & 30/43\\
R$_{in}$(km) & 26$\pm$1 & 48$\pm$2 & 55$\pm$2  & 40$\pm$5 & 49$\pm$2 & 54$\pm$2    & 48$\pm$2 \\
simpl*disk flux & 3.08$\pm$0.18	& 12.02$\pm$0.11		& 14.27$\pm$0.20	& 6.84$\pm$0.05	& 8.40$\pm$0.07	& 13.58$\pm$0.29 & 10.40$\pm$0.11\\
power-law flux 	& 2.92$\pm$0.09	& 5.12$\pm$0.06		& 7.14$\pm$0.08	& 3.66$\pm$0.12	& 7.43$\pm$0.21	&  8.22$\pm$0.06 & 3.03$\pm$0.51 \\
\hline
$\nu$\footnote{QPO centriod frequency} / Q$ = \nu / \delta \nu$\footnote{Quality factor}  &  5.60$\pm$0.06 / 2.10  & 5.87$\pm$0.07 / 2.67     & 5.54$\pm$0.05 / 3.14  & 5.51$\pm$0.09 / 3.15  & 5.35$\pm$0.07 / 3.01 &  5.60$\pm$0.06 / 2.10  & 4.56$\pm$0.20 / 4.40 \\
\hline
   &   &  &  &  model 3  &   &   &  \\
\hline
 $kT_{in} (keV)$ & 1.27$\pm$0.01  & 1.29$\pm$0.01      & 1.23$\pm$0.01   & 1.16$\pm$0.01   & 1.16$\pm$0.01  & 1.23$\pm$0.01  &  1.25$\pm$0.01 	\\
 N$_{diskbb}$	&  110$\pm$12 &  355$\pm$29 & 600$\pm$37 &  397$\pm$53 &  620$\pm$19 &  512$\pm$43 &   419$\pm$26  	\\
 $\Gamma$	& 2.54$\pm$0.10   & 2.39$\pm$0.12   & 2.40$\pm$0.13   & 2.47$\pm$0.23   & 2.32$\pm$0.01 & 2.39$\pm$0.12   &  2.43$\pm$0.17 		\\
 $E_{cut} (keV)$& 47.53$\pm$18.28 & 33.46$\pm$9.92  & 64$\pm$11 & 60.88$\pm$14.23 & 74$\pm$13& 63.02$\pm$19.24 & 38.11$\pm$18.33 \\
 $N_{cutoffpl}$	& 3.98$\pm$0.70   & 7.83$\pm$1.66    & 8.36$\pm$2.13   & 4.96$\pm$2.08   & 5.08$\pm$0.34& 9.39$\pm$2.06   &  4.79$\pm$1.42  \\
 $\chi^{2}$ / dof& 44/43          & 42/43  	    & 35/43 		& 69/41           & 37/44& 30/43          &  34/43\\
 R$_{in}(km)$	& 21$\pm$1  & 38$\pm$2  & 49$\pm$2  & 40$\pm$3  & 50$\pm$1& 45$\pm$2  &  41$\pm$1\\ 
 diskbb flux	& 2.28$\pm$0.12 	  & 8.09$\pm$0.09  	& 10.73$\pm$0.22  & 5.13$\pm$0.08   & 8.04$\pm$0.05  	& 9.22$\pm$0.08	 & 8.16$\pm$0.09  \\ 
 cutoffpl flux	& 3.74$\pm$0.08  	  & 9.08$\pm$0.36  	& 10.69$\pm$0.12 	& 5.52$\pm$0.27 & 7.77$\pm$0.31 & 12.30$\pm$0.35 & 5.36$\pm$0.19  \\ 
\hline
\hline
sections    & 80146-01-27-00  & 80146-01-49-00  & 80146-01-51-00 & 80146-01-56-00 & 80146-01-62-00 & 80146-01-65-00 \\
\hline
 &  &  & model 2 & &   \\ 
\hline

$\Gamma_{simpl}$    & 2.84$\pm$0.42	  & 2.69$\pm$0.28     & 2.58$\pm$0.20    & 2.59$\pm$0.22    & 2.30$\pm$0.18    & 2.67$\pm$0.42\\
f$_{scat}$   	    & 0.13$\pm$0.04	  & 0.17$\pm$0.04     & 0.04$\pm$0.01	 & 0.04$\pm$0.01    & 0.08$\pm$0.01   & 0.04$\pm$0.01\\
$kT_{in}$ ($keV$)   & 1.22$\pm$0.02	  & 1.18$\pm$0.02     & 1.17$\pm$0.01    & 1.18$\pm$0.01     & 1.15$\pm$0.01   & 1.16$\pm$0.01\\
N$_{diskbb}$      & 811$\pm$59  &  815$\pm$72  &  715$\pm$38 &  753$\pm$40 &  813$\pm$62 &  766$\pm$42 \\
$\Gamma$  	    & 2.88$\pm$0.15	  & 2.70$\pm$0.21     & 2.63$\pm$0.16	 & 2.61$\pm$0.14     & 3.18$\pm$0.19	& 2.56$\pm$0.15\\
N$_{pl}$   	    & 8.85$\pm$3.62	  & 7.93$\pm$2.92     & 4.88$\pm$1.64	 & 5.89$\pm$1.73     & 8.54$\pm$3.74	& 4.87$\pm$1.65\\
$\chi^{2}$ / dof  	      & 34/43		  & 26/43	      & 31/43		 & 36/43	     & 29/43		& 30/43\\
R$_{in}$(km) 	 & 57$\pm$2    & 57$\pm$3    & 54$\pm$2 	 & 55$\pm$1    & 57$\pm$2	& 55$\pm$2\\
simpl*disk flux & 16.87$\pm$0.45 & 15.78$\pm$0.22 & 10.45$\pm$0.20 & 11.59$\pm$0.10 & 12.07$\pm$0.13 & 10.54$\pm$0.08	\\
power-law flux 	 & 5.22$\pm$0.30  & 6.45$\pm$0.20  & 4.57$\pm$0.08  & 5.74$\pm$0.06  & 2.90$\pm$0.09 & 5.21$\pm$0.14\\
\hline
$\nu$ / Q$( = \nu / \delta \nu)$  & 5.29$\pm$0.03 / 4.01& 5.36$\pm$0.06 \ 2.08 & 4.49 $\pm$ 0.11 \ 3.95 & 4.83 $\pm$ 0.03 \ 6.15 &  4.46 $\pm$ 0.07 \ 3.67 & 4.94$\pm$0.05 \ 4.95  \\
\hline
   &   &  &  model 3    & &   \\
\hline
 $kT_{in} (keV)$   & 1.26$\pm$0.01   & 1.24$\pm$0.01	 & 1.18$\pm$0.01   & 1.19$\pm$0.01   & 1.15$\pm$0.01   & 1.17$\pm$0.01 \\
 N$_{diskbb}$	&  607$\pm$38 & 546$\pm$37 & 666$\pm$28 &  701$\pm$33 &  778$\pm$17 &  698$\pm$31 \\
 $\Gamma$	  & 2.40$\pm$0.14   & 2.29$\pm$0.12	 & 2.46$\pm$0.15   & 2.45$\pm$0.14   & 2.23$\pm$0.03   & 2.44$\pm$0.13\\
 $E_{cut} (keV)$ & 40.16$\pm$15.92 & 44.35$\pm$16.52 & 113$\pm$25& 114$\pm$24& 50.00      & 124$\pm$27\\
 $N_{cutoffpl}$	 & 8.40$\pm$2.15   & 8.16$\pm$1.78	 & 4.69$\pm$1.19   & 5.64$\pm$1.46   & 2.97$\pm$0.25   & 5.11$\pm$1.12\\
 $\chi^{2}$ / dof	     &  32/43	       & 21/43		 & 30/43           & 35/43           & 39/44           & 30/43\\
 R$_{in}(km)$	  & 49$\pm$2  & 47$\pm$2	 & 52$\pm$1  & 53$\pm$1  & 55$\pm$1  & 53$\pm$2\\ 
 diskbb flux	& 12.08$\pm$0.12 & 9.95$\pm$0.09 & 9.40$\pm$0.11 & 10.19$\pm$0.16 & 9.82$\pm$0.09 & 9.29$\pm$0.12\\ 
 cutoffpl flux	& 9.98$\pm$0.23  & 12.33$\pm$0.30 & 5.64$\pm$0.18 & 6.93$\pm$0.23 & 5.11$\pm$0.15 & 6.47$\pm$0.22 \\  
\hline

\end{tabular}
\end{minipage}
\end{table}

\clearpage
\begin{landscape}
\begin{table}
\begin{minipage}[t]{\columnwidth}
\scriptsize
\caption{ PCA$+$HEXTE spectral parameters of type-C \& type-B QPO observations. Fluxes are in units 10$^{-9}$ erg cm$^{-2}$ s$^{-1}$.} 
\label{ }
\centering
\begin{tabular}{ccccccccc}
\hline
\hline
&&&& type-C QPOs&&&&\\
sections & 80146-01-02-00 & 80146-01-03-00 & 80146-01-29-00 & 80146-01-30-00 & 80146-01-31-00 & 80146-01-32-00 &  80146-01-33-01 &  80146-01-33-00 \\
\hline

 &&&& model 1 &&&& \\

\hline
$kT_{in}$ ($keV$)& 0.67$\pm$0.04      & 0.71$\pm$0.03 	  & 0.75$\pm$0.03  & 0.69$\pm$0.03 & 0.74$\pm$0.03  & 0.75$\pm$0.03 & 0.76$\pm$0.04     & 0.78$\pm$0.04 \\
N$_{diskbb}$	 & 1075$\pm$278 & 953$\pm$205 & 3272$\pm$713 &  3231$\pm$847 & 1004$\pm$223 &  2252$\pm$489 &  2706$\pm$605 &  2556$\pm$557 \\
kT$_{0}$ 	 & $= kT_{in}$ 	      & $= kT_{in}$ 	 & $= kT_{in}$      & $= kT_{in}$  & $= kT_{in}$  & $= kT_{in}$ & $= kT_{in}$    & $= kT_{in}$\\
kT$_{e}$($keV$)  & 26.96$\pm$7.22    & 10.25$\pm$1.21	 & 17.11$\pm$2.79  & 16.26$\pm$2.10 & 19.46$\pm$4.11 & 16.12$\pm$2.14  & 24.88$\pm$7.61   & 24.37$\pm$6.74\\
$\tau$ 		 & 1.59$\pm$0.41     & 4.34$\pm$0.45 	& 2.73$\pm$0.48    & 2.99$\pm$0.36  & 2.43$\pm$0.50  & 3.04$\pm$0.38 & 1.87$\pm$0.68     & 1.91$\pm$0.64\\
N$_{comptt}$	 & 0.04$\pm$0.01    & 0.08$\pm$0.01	 & 0.19$\pm$0.03   & 0.19$\pm$0.02  & 0.05$\pm$0.01  & 0.16$\pm$0.02 & 0.12$\pm$0.03     & 0.12$\pm$0.03\\
             
diskbb flux	& 0.46$\pm$0.10  	   & 0.60$\pm$0.15 		& 2.87$\pm$0.09    & 1.61$\pm$0.12       & 0.81$\pm$0.12      & 2.06$\pm$0.18    & 2.71$\pm$0.08          & 3.00$\pm$0.09\\
CompTT flux	& 5.59$\pm$0.08  	   & 4.79$\pm$0.09 	  &  21.30$\pm$0.80  	   & 19.79$\pm$0.62     & 6.23$\pm$0.20     & 18.08$\pm$0.55 &  20.96$\pm$0.49	 &  21.33$\pm$0.33 \\
$\chi^{2}$ / dof & 63/72  	    & 93/72 	&  45/73  		  &  86/72	    & 57/72 	   &  75/72    & 57/73		 & 70/73 \\

R$_{in}$(km)	& 66$\pm$9    & 62$\pm$7     & 115$\pm$12    & 114$\pm$15 & 64$\pm$7 & 95$\pm$10  & 104$\pm$12	& 101$\pm$11  \\
\hline
&&&& type-B QPOs &&&&\\
sections & 80146-01-11-00 &  80146-01-12-00 &80146-01-16-00 &80146-01-27-00 &80146-01-48-00 &80146-01-49-00 & 80144-01-01-00 &  80135-02-02-00  \\
\hline
  & &&&  model 2  &&&&   \\

\hline
$\Gamma_{simpl}$ & 3.11$\pm$0.31   & 3.07$\pm$0.08  & 3.03$\pm$0.13  & 2.99$\pm$0.08 & 2.77$\pm$0.07 & 2.72$\pm$0.05 & 2.70$\pm$0.09 &  2.90$\pm$0.14		   \\
f$_{scat}$	 & 0.19$\pm$0.07   & 0.19$\pm$0.02  & 0.10$\pm$0.01  & 0.14$\pm$0.01 & 0.16$\pm$0.01 & 0.19$\pm$0.01 & 0.16$\pm$0.01 &  0.10$\pm$0.01		 	  \\
$kT_{in}$ ($keV$)& 1.20$\pm$0.05   & 1.22$\pm$0.02  & 1.22$\pm$0.02  & 1.20$\pm$0.02 & 1.18$\pm$0.02 & 1.17$\pm$0.02 & 1.18$\pm$0.02 &  1.20$\pm$0.02	  	   \\
N$_{diskbb}$	 &  173$\pm$68 &  566$\pm$52 &  532$\pm$41 &  877$\pm$59 & 829$\pm$63  &  891$\pm$70 &  757$\pm$70 &  767$\pm$64	    \\
 $\Gamma$ 	 & 2.83$\pm$0.05   & 2.82$\pm$0.05  & 2.80$\pm$0.07  & 2.73$\pm$0.05 & 2.65$\pm$0.04 & 2.69$\pm$0.03 & 2.68$\pm$0.04 &  2.60$\pm$0.04 	  	\\
 $N_{pl}$	 & 3.92$\pm$1.47   & 7.06$\pm$1.03  & 4.43$\pm$0.97  & 6.20$\pm$0.90 & 5.78$\pm$0.78 & 6.57$\pm$0.75 & 8.31$\pm$1.13 &  7.55$\pm$1.04		  	\\
$\chi^{2}$ / dof & 54/73 	  & 49/74	   &  51/74   	     &  49/74	 &  51/74  & 46/74 	 & 38/74 &  66/74		  	 	\\
R$_{in}$	 & 26$\pm$5 & 48$\pm$2  & 46$\pm$2 & 59$\pm$2 & 58$\pm$2 & 60$\pm$2 & 55$\pm$3 & 56$\pm$2 \\
simpl*disk flux  & 3.65$\pm$0.11 	& 13.26$\pm$0.21    & 10.98$\pm$0.32 	    & 18.21$\pm$0.21 & 16.39$\pm$0.37 &  18.30$\pm$0.33 & 15.63$\pm$0.26 & 14.61$\pm$0.30      \\
powerlaw flux 	 &  3.84$\pm$0.19   & 7.05$\pm$0.18	  & 4.60$\pm$0.21       & 7.57$\pm$0.24 & 8.68$\pm$0.20 & 8.95$\pm$0.15 & 11.59$\pm$0.22 & 12.79$\pm$0.29 \\
\hline
sections  & 80144-01-01-02 & 80146-01-51-00 & 80146-01-56-00 & 80146-01-62-00 & 80146-01-65-00 & 80146-01-68-00  \\
\hline
$\Gamma_{simpl}$ & 3.06$\pm$0.17 & 2.92$\pm$0.30 & 2.75$\pm$0.17 & 4.00$\pm$0.96 & 2.73$\pm$0.27 & 3.11$\pm$0.55	 \\
f$_{scat}$	 & 0.11$\pm$0.02 & 0.05$\pm$0.02 & 0.06$\pm$0.01 & 0.04$\pm$0.02 & 0.04$\pm$0.01 & 0.04$\pm$0.02	  \\
$kT_{in}$ ($keV$)& 1.12$\pm$0.02 & 1.16$\pm$0.02 & 1.17$\pm$0.01 & 1.14$\pm$0.02 & 1.15$\pm$0.01 & 1.15$\pm$0.02	   \\
N$_{diskbb}$	 &  533$\pm$62 &  759$\pm$59 &  824$\pm$53 &  818$\pm$72 &  775$\pm$58 &  651$\pm$66  \\
 $\Gamma$ 	 & 2.64$\pm$0.05 & 2.54$\pm$0.08 & 2.55$\pm$0.05 & 2.44$\pm$0.05 & 2.55$\pm$0.06 & 2.50$\pm$0.04	\\
 $N_{pl}$	 & 4.13$\pm$0.65 & 3.74$\pm$0.96 & 4.41$\pm$0.75 & 3.60$\pm$0.62 & 4.70$\pm$0.89 & 6.06$\pm$0.80	\\
$\chi^{2}$ / dof &   60/71   &  52/74       &   45/74             & 35/73   &   47/74     & 34/74	 	\\
R$_{in}$	 & 46$\pm$3 & 55$\pm$2 & 57$\pm$2 & 57$\pm$2 & 56$\pm$2 &  51$\pm$3    \\
simpl*disk flux &  7.13$\pm$0.15 &  11.14$\pm$0.28 &  13.11$\pm$0.31 & 10.28$\pm$0.34   & 10.96$\pm$0.19 & 8.65$\pm$0.26 \\
powerlaw flux 	 & 6.34$\pm$0.11 & 7.40$\pm$0.17 & 8.27$\pm$0.20 & 8.97$\pm$0.18   & 9.05$\pm$0.13 & 13.27$\pm$0.22 \\
\hline
\end{tabular}
\end{minipage}
\end{table}
\end{landscape}
\begin{table}
\begin{minipage}[t]{\columnwidth}
\scriptsize
\caption{Spectral parameters of ObsIDs 80146-01-47-00 (type-C QPO) \& 80146-01-48-00 (type-B QPO). Fluxes are in the unit of 10$^{-9}$ erg cm$^{-2}$ s$^{-1}$ }. 
\label{ }
\centering 
\begin{tabular}{c|c|c|c|c}
\hline
  & & 80146-01-47-00 & & 80146-01-48-00  \\
\hline
parameter	&  (a)	& (b)		& (c) 	&  (d) 	\\
\hline
&	& model 1	&	&   \\
\hline
 $kT_{in}$ ($keV$) &	1.06$\pm$0.05		& 1.08$\pm$0.03		& 1.13$\pm$0.06	& 1.23$\pm$0.01	\\
N$_{diskbb}$	  &	 383$\pm$97	& 421$\pm$74	& 357$\pm$54 &  546$\pm$24	\\
kT$_{0}$	& 	0.45		&  0.25			& 0.41$\pm$0.20		& 0.40	\\
kT$_{e}$	&	15.30		& 8.90$\pm$2.80		& 10.26$\pm$3.38	& 11.21$\pm$5.12 \\
$\tau$		& 	3.35$\pm$0.08		& 5.19$\pm$1.00		& 4.68$\pm$1.13	& 4.26$\pm$1.24	\\
N$_{comptt}$	&	0.31$\pm$0.02		& 1.02$\pm$0.13		& 0.51$\pm$0.36	& 0.45$\pm$0.12	\\
diskbb flux	& 	3.03$\pm$0.21		&  3.57$\pm$0.12		& 3.91$\pm$0.09	 & 10.01$\pm$0.19  \\
comptt flux	& 	11.35$\pm$0.06		&   11.16$\pm$0.09	& 11.52$\pm$0.19	  & 10.43$\pm$0.04  \\
$\chi^{2}$ / dof& 	37/42		& 39/41			& 43/40	  & 33/43 \\
R$_{in}$	& 	39$\pm$5		&  41$\pm$4	&  38$\pm$3 & 47$\pm$2  \\
\hline
$\nu$ / Q 	& 6.47$\pm$0.06 / 5.30$\pm$0.91  & 7.06$\pm$0.06 / 6.85$\pm$0.80	& 7.35$\pm$0.03 / 5.49$\pm$1.10 & 5.45$\pm$0.15 / 3.55$\pm$0.75   \\
rms(\%)		& 11.64$\pm$0.13 		& 9.88$\pm$0.12			& 9.60$\pm$0.18			&  8.25$\pm$0.13\\
\hline
&&model 1 with kT$_{0}$=kT$_{in}$ &&\\ 
\hline
 $kT_{in}$ ($keV$)&   0.85$\pm$0.03            & 0.85$\pm$0.03            & 0.88$\pm$0.02          & --\\
N$_{diskbb}$&    1735$\pm$258 &  1702$\pm$283 &  1595$\pm$216 & -- \\
kT$_{0}$&    kT$_{0}$$=kT_{in}$        & $=kT_{in}$       & $=kT_{in}$        & --\\
   kT$_{e}$ & 15.35                 & 12.71$\pm$5.58        & 14.18$\pm$5.91        & --\\
$\tau$		& 3.30$\pm$0.09            & 3.75$\pm$1.88         & 3.45$\pm$1.32            & --\\
N$_{comptt}$&  0.14$\pm$0.01           & 0.17$\pm$0.07        & 0.15$\pm$0.05              &  --\\
$\chi^{2}$ / dof& 38/42                & 40/41                & 47/41                & --\\
diskbb flux	&3.56$\pm$0.12 & 3.61$\pm$0.15           & 4.11$\pm$0.11           & --\\
comptt flux	&10.72$\pm$0.09& 11.03$\pm$0.20        & 11.18$\pm$0.09           & --\\
R$_{in}$	& 83$\pm$6        & 83$\pm$7                & 80$\pm$5                & --\\

\hline
&	& model 2 without power-law &	&   \\
\hline
$\Gamma_{simpl}$	& 2.61$\pm$0.03	& 2.65$\pm$0.02 	& 2.65$\pm$0.01	& 2.76$\pm$0.03	 \\
f$_{scat}$		& 0.44$\pm$0.02	& 0.44$\pm$0.01		& 0.43$\pm$0.01	& 0.24$\pm$0.01  \\
$kT_{in}$ ($keV$)		& 0.85$\pm$0.03	& 0.87$\pm$0.03 & 0.88$\pm$0.02	& 1.10$\pm$0.02  \\
N$_{diskbb}$		&  2659$\pm$493	&  2523$\pm$482 &  2372$\pm$299 &  1395$\pm$116	  \\
$\chi^{2}$ / dof	& 37 / 42	& 47 / 42		& 58/42		& 37/42		\\
R$_{in}$	& 103$\pm$10	&  101$\pm$9	& 98$\pm$6 & 75$\pm$3    \\
simpl*disk flux & 14.34$\pm$0.10	&  14.67$\pm$0.12		& 15.34$\pm$0.09	& 20.21$\pm$0.20 	\\
 
\hline
  & & model 3 & &  \\  
\hline
 $kT_{in} (keV)$ & 1.10$\pm$0.03 & 1.06$\pm$0.03   & 1.10$\pm$0.02   & 1.23$\pm$0.01\\
 N$_{diskbb}$	&  285$\pm$36 & 484$\pm$98 &  424$\pm$50 &  558$\pm$35 \\
 $\Gamma$	& 2.43$\pm$0.03  & 2.10$\pm$0.20   & 2.18$\pm$0.11   & 2.28$\pm$0.13	\\
 $E_{cut} (keV)$& 106.07	 & 28.59$\pm$12.21 & 35.36$\pm$9.66  & 42.15$\pm$16.62\\
 $N_{cutoffpl}$	& 9.12$\pm$0.82  & 5.57$\pm$1.80   & 6.35$\pm$1.1.24 & 6.89$\pm$1.64	\\
 $\chi^{2}$ / dof& 36/43	 & 39/42	   & 44/42	     & 33/43 \\
 R$_{in}(km)$	& 34$\pm$2 & 44$\pm$5  & 41$\pm$2  & 47$\pm$2 \\ 
 diskbb flux	& 2.78$\pm$0.20	& 3.73$\pm$0.12    & 3.97$\pm$0.07 	     & 9.95$\pm$0.12 \\ 
 cutoffpl flux	& 11.68$\pm$0.05  & 10.98$\pm$0.09	   & 11.43$\pm$0.12	& 10.49$\pm$0.04\\ 
\hline
\end{tabular}
\end{minipage}
\end{table}

\clearpage

\begin{table}
\begin{minipage}[t]{\columnwidth}
\scriptsize
\caption{PCA$+$HEXTE spectral parameters of observations 80146-01-47-00 \& 80146-01-48-00. Fluxes are in the unit 10$^{-9}$ erg cm$^{-2}$ s$^{-1}$. }
\label{ }
\centering
\begin{tabular}{cccc}
\hline
\hline
sections &  80146-01-47-00 (a)  & (c)  &  80146-01-48-00 (d) \\
\hline
  & &model 1 & \\
\hline
$kT_{in}$ ($keV$)& 0.91$\pm$0.04     & 0.92$\pm$0.02     & -- \\
N$_{diskbb}$	 &  1239$\pm$279 &  1288$\pm$146 & -- \\
kT$_{0}$ 	 & = $kT_{in}$    & = $kT_{in}$     & --\\

kT$_{e}$($keV$)  & 31.06$\pm$0.79     & 39.96$\pm$15.83   & --\\
$\tau$ 		 & 1.62$\pm$0.05     & 1.18$\pm$0.50     & --\\
N$_{comptt}$	& 0.06$\pm$0.01      & 0.05$\pm$0.02     & --\\
diskbb flux	&  3.80$\pm$0.12 	     & 4.19$\pm$0.11         & --\\
CompTT flux	& 18.53$\pm$0.32	     & 19.32$\pm$0.28	 & --\\
$\chi^{2}$ / dof  &  60/75  		     & 73/75		 & -- \\

R$_{in}$(km)	& 71$\pm$8     & 72$\pm$4		& --  \\

\hline
  &  &  model 2   	&   \\
\hline
$\Gamma_{simpl}$ & 2.36$\pm$0.17	& 2.60$\pm$0.03		& 2.77$\pm$0.06	 	 \\
f$_{scat}$	 & 0.28$\pm$0.03	& 0.39$\pm$0.02		& 0.16$\pm$0.01  \\
$kT_{in}$ ($keV$)& 1.08$\pm$0.06	& 0.99$\pm$0.03 	& 1.18$\pm$0.02	   \\
N$_{diskbb}$	 &  454$\pm$173	&  1157$\pm$163 	&  829$\pm$63  	  \\
 $\Gamma$ 	 & 2.75$\pm$0.09	& 2.82$\pm$0.04 	& 2.64$\pm$0.04	\\
 $N_{pl}$	 & 10.17$\pm$2.81	& 4.38$\pm$0.52	        & 5.78$\pm$0.78	\\
$\chi^{2}$ / dof & 55/74		& 76 / 74		& 51/74	 	\\
R$_{in}$	 &   43$\pm$8 	&   68$\pm$5 	&   58$\pm$2     \\
simpl*disk flux  &  8.37$\pm$0.10		&  14.71$\pm$0.32		&  16.39$\pm$22  \\
powerlaw flux 	 &  11.85$\pm$0.21		&  4.29$\pm$0.20		&  8.68$\pm$0.08	 \\
\hline
 & &model 4, power-law for HEXTE only  & \\
\hline
 $\Gamma$ 	& 2.75$\pm$0.10 & 2.82$\pm$0.04 & 2.66$\pm$0.04	\\
 $N_{pl}$	& 16.44$\pm$5.50& 20.46$\pm$2.23& 11.46$\pm$1.45	\\
 $\chi^{2}$ / dof& 22/30	& 25/30    & 29/30 \\
  pl flux	&  3.47$\pm$0.08 & 3.44$\pm$0.09 & 3.32$\pm$0.12	\\ 
\hline
\end{tabular}
\end{minipage}
\end{table}

\clearpage

\nocite{*}
\bibliography{ref}{}
 \bibliographystyle{mnras}


\bsp	
\label{lastpage}
\end{document}